\def\me{Elisabeth Lobe}

\def\myorcid{0000-0002-3473-8906}
\def\myaffil{Institute for Software Technology, German Aerospace Center (DLR), Lilienthalplatz 7, 38108 Braunschweig, Germany}

\def\docsubject{Article}
\def\doctitle{Optimal Sufficient Requirements on the Embedded Ising Problem in Polynomial Time}
\def\dockeywords{Ising Problem, Embedding, Quantum Annealing, Linear Optimization}

\RequirePackage{ifthen}
    \provideboolean{journal}
    \providecommand{\ifjournalversion}[2]{%
        \ifthenelse{\boolean{journal}}{#1}{#2}%
    }
\setboolean{journal}{false}

\documentclass[
    parskip=half,
    numbers=noenddot,
    bibliography=totoc,
]
{scrartcl}

\RequirePackage{ifthen}

\usepackage{cmap}  %
\usepackage[utf8]{inputenc}
\usepackage[T1]{fontenc}
\usepackage[ngerman, english]{babel}

\usepackage[backend=biber, style=trad-abbrv]{biblatex}  %
\usepackage{csquotes}

\usepackage{geometry}   %
\makeatletter
\renewcommand{\subsubsection}{\@startsection%
    {subsubsection}%
    {3}%
    {0mm}%
    {0.3\baselineskip}%
    {0.3\baselineskip}%
    {\sffamily\bfseries}}%
\makeatother

\usepackage{authblk} %

\usepackage{chngcntr}
    \counterwithout{equation}{section}              %

\AtBeginDocument{

}

\usepackage{amssymb}
\usepackage{amsmath}
\usepackage{mathtools}
    \mathtoolsset{showonlyrefs=true}

\usepackage{amsthm}

\usepackage{thmtools}           % list of theorems
    \newtheorem{theorem}{Theorem}
    \newtheorem{corollary}[theorem]{Corollary}
    \newtheorem{lemma}[theorem]{Lemma}

    \theoremstyle{definition}
    \newtheorem{definition}[theorem]{Definition}

    \declaretheoremstyle[style=definition,
                         notebraces={}{},
                         name={},
                         numbered=no,
                         headindent=-0.444em,
                         notefont=\bfseries]{problemstyle}
    \theoremstyle{problemstyle}
    \declaretheorem[style=problemstyle]{prob*}

\usepackage{framed}
    \newenvironment{fprob*}[1][]
        {\begin{framed}\vspace{-2ex}\begin{prob*}[\normalfont #1]\phantomsection}
        {\end{prob*}\vspace{-2ex}\end{framed}}

\usepackage{etoolbox}                               % to have local enumeration of equations in proofs 
    \newcounter{equationstore}      
    \newcounter{proofnum}
    \AtBeginEnvironment{proof}{\setcounter{equationstore}{\value{equation}}
    \refstepcounter{proofnum}
    \setcounter{equation}{0}
    }
    \AtEndEnvironment{proof}{\setcounter{equation}{\value{equationstore}}}

\usepackage[dvipsnames, table]{xcolor}

\usepackage{xurl}
\usepackage[pdftitle={\doctitle},
            pdfauthor={\me},
            pdfkeywords={\dockeywords},
            pdfsubject={\docsubject},
            pdfdisplaydoctitle=true
           ]{hyperref}
\hypersetup{breaklinks=true}
                
\usepackage[shortlabels]{enumitem}
\usepackage{bbm}
\usepackage{pifont}

\usepackage{tikz}
    \usetikzlibrary{angles}
    \usetikzlibrary{shapes.geometric}

\usepackage{currfile}
\usepackage{orcidlink}

\newcommand{\R}{\mathbb{R}}
\newcommand{\N}{\mathbb{N}}
\newcommand{\Z}{\mathbb{Z}}
\newcommand{\Q}{\mathbb{Q}}
\newcommand{\X}{\mathbb{X}}
\newcommand{\Zero}{\mathbb{O}}

\newcommand{\One}{\mathbbm{1}}

\newcommand{\cupdot}{\mathbin{\mathaccent\cdot\cup}}

\newcommand{\closure}[2][3]{{}\mkern#1mu\overline{\mkern-#1mu#2}}

\DeclareMathOperator*{\argmin}{arg\,min}

\DeclareMathOperator{\sign}{sign}

\makeatletter
\providecommand*{\bigcupdot}{%
  \mathop{%
    \vphantom{\bigcup}%
    \mathpalette\@bigcupdot{}%
  }%
}
\newcommand*{\@bigcupdot}[2]{%
  \ooalign{%
    $\m@th#1\bigcup$\cr
    \sbox0{$#1\bigcup$}%
    \dimen@=\ht0 %
    \advance\dimen@ by -\dp0 %
    \sbox0{\scalebox{2}{$\m@th#1\cdot$}}%
    \advance\dimen@ by -\ht0 %
    \dimen@=.5\dimen@
    \hidewidth\raise\dimen@\box0\hidewidth
  }%
}
\makeatother

\renewcommand{\emph}{\textbf}

\def\clap#1{\hbox to 0pt{\hss#1\hss}}

\def\mathclap{\mathpalette\mathclapinternal}
    \def\mathclapinternal#1#2{\clap{$\mathsurround=0pt#1{#2}$}}

\newcommand{\refProb}[2][the\ ]{#1\textnormal{\textsc{\nameref{prob:#2}}}}
\newcommand{\refProbNamed}[2][the\ ]{#1\textnormal{\textsc{\ref{prob:#2}}}}

\makeatletter
\def\namedlabel#1#2{\begingroup
    \def\@currentlabel{#2}%
    \phantomsection\label{#1}\endgroup
}
\makeatother
 
\makeatletter
\def\provideenvironment{\@star@or@long\provide@environment}
\def\provide@environment#1{%
        \@ifundefined{#1}%
                {\def\reserved@a{\newenvironment{#1}}}%
                {\def\reserved@a{\renewenvironment{dummy@environ}}}%
        \reserved@a
}
\def\dummy@environ{}
\makeatother

    \provideenvironment{abstract}{\textbf{Abstract }}{}
    \provideenvironment{keywords}{\textbf{Keywords }}{}
    \provideenvironment{acknowledgements}{\textbf{Acknowledgments }}{}

\usepackage{currfile}       %

\usepackage{xifthen}        %
\usepackage{xstring}        %
\usepackage{etoolbox}       %

\usepackage{amsmath}
\usepackage{bm}

\usepackage[dvipsnames, table]{xcolor}

\usepackage{tikz}
    \usetikzlibrary{calc}
    \usetikzlibrary{fadings}
    \usetikzlibrary{backgrounds}
    \usetikzlibrary{external}
        \tikzexternaldisable
        \newenvironment{externalize}{}{}

\usepackage{pgfplots}
    \pgfplotsset{compat=1.16}
    \usepgfplotslibrary{fillbetween}
    \usepackage{pgfmath}
    \usepackage{pgfplotstable}

\usepackage{diagbox}        %

\definecolor{lightgreen}{RGB}{146,208,80}
\definecolor{lightblue}{RGB}{0,176,240}
\definecolor{darkgreen}{rgb}{0,0.39,0}
\definecolor{darkblue}{rgb}{0,0,0.5}
\definecolor{lightgreenblue}{RGB}{73,192,160}
\definecolor{darkgreenblue}{rgb}{0,0.195,0.25}

\def\rows{3}
\def\cols{3}
\def\shift{6}
\def\xshift{0}
\def\yshift{0}

\def\scale{0.2}
\def\LineWidthColored{9pt}
\def\lineDist{2.5}

\tikzstyle{node} = [circle, draw, fill=white, inner sep=0pt, text centered, minimum height=3ex*\scale]
\tikzstyle{big} = [circle, draw, fill=white, inner sep=0pt, text centered, minimum height=6ex*\scale]
\tikzstyle{nodec} = [circle, draw, fill=lightgreen, inner sep=0pt, text centered, minimum height=3.5ex*\scale]
\tikzstyle{broken} = [circle, fill=white, densely dashed, inner sep=0pt, text centered, minimum height=3ex*\scale]
\tikzstyle{recphantom} = [nodec, rectangle, minimum height=\lineDist*1ex*\scale, minimum width=\lineDist*1ex*\scale, white]

\tikzstyle{every text node part}=[font=\scriptsize]

\tikzstyle{hamiltonian} = [line width=\scale*\LineWidthColored, rounded corners=2pt, lightgreenblue]
\tikzstyle{edge} = []

\def\brokenQubitsColor{Gray}

\newboolean{drawBrokenQubits}
\setboolean{drawBrokenQubits}{true}

\newboolean{drawQubitsBipartite}
\setboolean{drawQubitsBipartite}{false}
\def\qubitsPartitionColorA{}
\def\qubitsPartitionColorB{black}

\makeatletter
\define@key{chimerakeys}{row}{\def\chimeraRow{#1}}
\define@key{chimerakeys}{col}{\def\chimeraCol{#1}}
\define@key{chimerakeys}{hov}{\def\chimeraHoV{#1}}
\define@key{chimerakeys}{dep}{\def\chimeraDep{#1}}
\tikzdeclarecoordinatesystem{chimera}%
{%
    \setkeys{chimerakeys}{#1}%
    \pgfpointadd{\pgfpointxy{\xshift}{\yshift}}{
    \pgfpointadd{\pgfpointscale{\shift}{\pgfpointadd{\pgfpointxy{\chimeraCol - 1}{1}}{\pgfpointscale{-1}{\pgfpointxy{0}{\chimeraRow}}}}}
                {\ifthenelse{\equal{\chimeraHoV}{h}}{\pgfpointxy{\chimeraDep*1.333 - 1.333}{0}}{\pgfpointxy{2}{3.333 - \chimeraDep*1.333}}}}
}

\newcommand{\chimeraCoord}[4]{(chimera cs:row=#1, col=#2, hov=#3, dep=#4)}

\def\brokenQubits{}

\newcommand{\setBrokenQubits}[1]{
    \def\brokenQubits{}
    \addBrokenQubits{#1}
}

\newcommand{\addBrokenQubits}[1]{
    \foreach \x in #1 {
        \listxadd\brokenQubits{\x}
    }
}

\newcommand{\ifIsBrokenQubit}[3]{
    \xifinlist{#1}{\brokenQubits}{#2}{#3}
}

\newcommand{\ifIsBrokenEdge}[4]{
    \ifboolexpr{ test {\xifinlist{#1}{\brokenQubits}} or test {\xifinlist{#2}{\brokenQubits}} }{#3}{#4}
}

\newcommand{\ifDrawBrokenQubits}[2]{
    \ifthenelse{\boolean{drawBrokenQubits}}{#1}{#2}
}

\newcommand{\ifDrawQubitsBipartite}[2]{
    \ifthenelse{\boolean{drawQubitsBipartite}}{#1}{#2}
}

\newcommand{\drawBrokenChimera}{

    \foreach \r in {1,...,\rows}{
        \foreach \c in {1,...,\cols}{

            \foreach \d in {1,...,4}{
                \drawQubit{\r}{\c}{h}{\d}
                \drawQubit{\r}{\c}{v}{\d}
            }

            \foreach \x in {1,...,4}{
                \foreach \y in {1,...,4}{
                    \drawEdgeIntra{\r}{\c}{\x}{\y}
            }}
    }}

    \foreach \r in {1,...,\rows} {
        \foreach \c in {2,...,\cols} {
            \foreach \x in {1,...,4}{
                \pgfmathparse{int(\c - 1)}
                \drawEdge{\r}{\pgfmathresult}{v}{\x}{\r}{\c}{v}{\x}
            }
    }}

    \foreach \c in {1,...,\cols} {
        \foreach \r in {2,...,\rows} {
            \foreach \x in {1,...,4}{
                \pgfmathparse{int(\r - 1)}
                \drawEdge{\pgfmathresult}{\c}{h}{\x}{\r}{\c}{h}{\x}
            }
    }}
}

\newcommand{\drawQubit}[4]{
    \ifIsBrokenQubit{(#1-#2-#3#4)}{
        \ifDrawBrokenQubits{
            \node[broken, draw=\brokenQubitsColor] (#1-#2-#3#4) at \chimeraCoord{#1}{#2}{#3}{#4} {};
        }{
            \node[node, opacity=0] (#1-#2-#3#4) at \chimeraCoord{#1}{#2}{#3}{#4} {};
        }
    }{
        \ifDrawQubitsBipartite{
            \ifnumodd{\r + \c}{
                \ifthenelse{\isin{v}{#3}}{
                    \node[node, fill=\qubitsPartitionColorA] (#1-#2-#3#4) at \chimeraCoord{#1}{#2}{#3}{#4} {};
                }{
                    \node[node, fill=\qubitsPartitionColorB] (#1-#2-#3#4) at \chimeraCoord{#1}{#2}{#3}{#4} {};
                }
            }{
                \ifthenelse{\isin{v}{#3}}{
                    \node[node, fill=\qubitsPartitionColorB] (#1-#2-#3#4) at \chimeraCoord{#1}{#2}{#3}{#4} {};
                }{
                    \node[node, fill=\qubitsPartitionColorA] (#1-#2-#3#4) at \chimeraCoord{#1}{#2}{#3}{#4} {};
                }
            }
        }{
            \node[node] (#1-#2-#3#4) at \chimeraCoord{#1}{#2}{#3}{#4} {};
        }
    }
}

\newcommand{\drawEdgeIntra}[4]{
    \drawEdge{#1}{#2}{h}{#3}{#1}{#2}{v}{#4}
}

\newcommand{\drawEdge}[8]{
    \ifIsBrokenEdge{(#1-#2-#3#4)}{(#5-#6-#7#8)}{
        \ifDrawBrokenQubits{
            \draw[\brokenQubitsColor, dashed] (#1-#2-#3#4) -- (#5-#6-#7#8);}{}
    }{
        \draw[edge] (#1-#2-#3#4) -- (#5-#6-#7#8);
    }
}

\newcommand{\drawColoredQubit}[2]{
    \node[nodec, fill=#2] at #1 {};
}

\newcommand{\drawColoredEdge}[3]{
    \draw[line width=\scale*\LineWidthColored, #3] #1 -- #2;
}

\newcommand{\drawColoredEdges}[2]{
    \pgfmathsetmacro{\qubitbefore}{}
    \foreach \x in {#1} {
        \drawColoredEdge{\qubitbefore}{\x}{#2}
        \global\let\qubitbefore=\x
    }
    \foreach \x in {#1} {
        \drawColoredQubit{\x}{#2}
    }
}

\newcommand{\drawHorizontalFromTo}[5]{
    \foreach \c in {#3,...,#4} {
        \ifthenelse{\c < #4}{
            \pgfmathtruncatemacro\colBefore{\c + 1}
            \drawColoredEdge{(#1-\colBefore-v#2)}{(#1-\c-v#2)}{#5}
        }{}
        \drawColoredQubit{(#1-\c-v#2)}{#5}
    }
}

\newcommand{\drawVerticalFromTo}[5]{
    \foreach \r in {#3,...,#4} {
        \ifthenelse{\r < #4}{
            \pgfmathtruncatemacro\rowBefore{\r + 1}
            \drawColoredEdge{(\rowBefore-#1-h#2)}{(\r-#1-h#2)}{#5}
        }{}
        \drawColoredQubit{(\r-#1-h#2)}{#5}
    }
}

\newcommand{\drawCrossFromTo}[9]{

    \drawColoredEdge{(#1-#2-v#3)}{(#1-#2-h#4)}{#9}
    \drawHorizontalFromTo{#1}{#3}{#5}{#6}{#9}
    \drawVerticalFromTo{#2}{#4}{#7}{#8}{#9}
}

\newcommand{\setcolor}[1]{
    \IfEqCase{#1}{%
        {1}{\colorlet{currentcolor}{lightblue}}%
        {2}{\colorlet{currentcolor}{NavyBlue}}%
        {3}{\colorlet{currentcolor}{Blue}}%
        {4}{\colorlet{currentcolor}{darkblue}}%
        {5}{\colorlet{currentcolor}{darkgreen}}%
        {6}{\colorlet{currentcolor}{ForestGreen}}%
        {7}{\colorlet{currentcolor}{lightgreen}}%
        {8}{\colorlet{currentcolor}{GreenYellow}}%
        {9}{\colorlet{currentcolor}{orange}}%
        {10}{\colorlet{currentcolor}{red}}%
        {11}{\colorlet{currentcolor}{purple}}%
        {12}{\colorlet{currentcolor}{violet}}%
    }[\colorlet{currentcolor}{#1}]%
}

\newcounter{colorCounter}
\newcommand{\setcolorCounter}{
    \setcolor{\thecolorCounter}
    \stepcounter{colorCounter}
    \ifthenelse{\thecolorCounter > 12}{\setcounter{colorCounter}{1}}{}
}

\newcommand{\drawCrossFromToCC}[8]{
    \setcolorCounter
    \drawCrossFromTo{#1}{#2}{#3}{#4}{#5}{#6}{#7}{#8}{currentcolor}
}

\newcommand{\setOppositeDirection}[1]{
    \IfEqCase{#1}{%
        {north}{\def\oppositeDirection{south}}%
        {south}{\def\oppositeDirection{north}}%
        {west}{\def\oppositeDirection{east}}%
        {east}{\def\oppositeDirection{west}}%
    }%
}

\newcommand{\inputFromHere}[1]{%
    \edef\pathhere{\currfiledir}%
    
    \IfEndWith*{\pathhere}{/}{%
        \input{\pathhere#1}%
    }{%
        \input{\pathhere/#1}%
    }%

}

\newcommand{\parsedotsaway}[1]{%
    \StrCut{#1}{../}{\leftofdots}{\rightofdots}%
    \StrCount{\leftofdots}{/}[\numberofslashes]%
    \pgfmathparse{int(\numberofslashes - 1)}%
    \StrCut[\pgfmathresult]{\leftofdots}{/}{\leftofslash}{\rightofslash}%
    \edef\parsedaway{\leftofslash/\rightofdots}%
    \IfSubStr{\parsedaway}{..}{%
        \parsedotsaway{\parsedaway}%
    }{}%
}

\newcommand{\vertices}{{\closure{V}}}
\newcommand{\edges}{{\closure{E}}}
\newcommand{\outerneighbors}{{\closure{N}}}
\newcommand{\outeredges}{{\closure{D}}}
\newcommand{\cutpoly}{\Theta}
\newcommand{\totalweight}{\lambda}

\newcommand{\varthetaHalf}{\vartheta_{\!\text{\tiny$\frac{1}{2}$}}}

\newcommand{\fC}{\mathfrak{C}}

\usepackage[backend=biber, style=trad-abbrv]{biblatex}  %
\begin{document}

\newbox{\orcidboxme}    % workaround since orcidlink does not work directly in \author
\sbox{\orcidboxme}{\orcidlink{\myorcid}}
\newcommand{\orcidme}{\usebox{\orcidboxme}}

\title{\doctitle}

\author[1]{\me~\orcidme}
\author[2]{Volker Kaibel}

% \authorrunning{Short form of author list} % if too long for running head
 
\affil[1]{\myaffil
       % Tel.: \mytel \\
       % \email{\mymail}
      }

\affil[2]{\mbox{Institute for Mathematical Optimization},~\mbox{Otto-von-Guericke Universit{\"a}t} Magdeburg, Universit{\"a}tsplatz 2, 39106 Magdeburg, Germany
       % Tel.: \mytel \\
       % \email{\mymail}
      }
      
\date{\normalsize 08.02.2023}

\phantomsection\pdfbookmark[section]{Title}{title}
{\centering
\begin{minipage}{0.75\textwidth} 
    \maketitle
\end{minipage}\hfill\\[1.0\baselineskip]}

\renewenvironment{abstract}{\textbf{Abstract }}{}
\phantomsection\pdfbookmark[subsection]{Abstract}{abstract}
\begin{abstract}
    One of the central applications for quantum annealers is to find the solutions of Ising problems.
Suitable Ising problems, however, need to be formulated such that they, on the one hand, respect the specific restrictions of the hardware 
and, on the other hand, represent the original problems which shall actually be solved. 
We evaluate sufficient requirements on such an embedded Ising problem analytically and transform them into a linear optimization problem. 
With an objective function aiming to minimize the maximal absolute problem parameter, 
the precision issues of the annealers are addressed.
Due to the redundancy of several constraints, we can show that the formally exponentially large optimization problem can be reduced
and finally solved in polynomial time for the standard embedding setting where the embedded vertices induce trees. 
This allows to formulate provably equivalent embedded Ising problems in a practical setup. 

\end{abstract}

\phantomsection\pdfbookmark[subsection]{Keywords}{keywords}

% \begin{acknowledgements}
%     \input{adds/acknowledgments}
% \end{acknowledgements}

\section{Introduction}

\subsection{Background}

The interest in quantum annealers, such as the devices developed by the company D-Wave Systems Inc., 
is still undiminished due to their ongoing fast progression. 
By implementing the adiabatic evolution of an \emph{Ising problem} over qubits formed by overlapping superconducting loops, they promise to solve NP-hard problems.
Although several physical effects prevent the ideal realization of the underlying adiabatic theorem, and optimal solutions can thus only be found with some probability, 
the experimental results appear to be promising for certain applications~\cite{stollenwerk2021agile}.  
However, the advantage over classical computation is still under discussion~\cite{junger2021quantum}. 

`Programming' such an annealer means to provide the input parameters of the specific implemented Ising problem, 
that is, the \emph{weights} on the vertices and the \emph{strengths} on the edges of a specific hardware graph. 
The \emph{Chimera} and \emph{Pegasus} hardware architectures are currently available~\cite{boothby2020next} 
and a new one, called \emph{Zephyr}, was recently announced but is not yet released~\cite{boothby2021zephyr}.
Interesting applications, however, do usually not match those graphs straightforwardly but require what is known as an \emph{embedding}~\cite{choi2011minor}, 
where each vertex of the original problem is mapped to several vertices in the hardware graph to represent the desired connectivity. 
Unfortunately, the problem of finding such an embedding is itself an NP-hard problem~\cite{lobe2021minor}. 
Although the connectivity is increased with every new hardware release, it is apparent that all of the graphs yield some kind of locality due to physical restrictions.
Therefore, the development of a completely connected hardware graph in the future is rather unlikely and the embedding problem will remain relevant in the long term.  
In order to circumvent this bottleneck and nevertheless enable experiments on these machines for the users, 
precalculated and generally applicable embedding templates provide a good starting point, such as for the complete graph~\cite{lobe2021embedding}.
Furthermore, the D-Wave API provides heuristic algorithms in the package \texttt{minorminor}~\cite{dwave2022minorminor}, 
which are mainly based on an implementation of~\cite{cai2014practical}.
 
However, with only the embedding, we still cannot perform calculations on the D-Wave machine. 
We need to bring together the two different problems: the original one that shall be solved and the one that can be solved with the annealer. 
That is, we need to find suitable parameters, the weights and strengths, of an Ising problem working on the hardware subgraph induced by the embedding.
The resulting \emph{embedded Ising problem} should represent the original Ising problem 
such that the corresponding solutions can be retrieved from the output of the quantum annealer (at least in theory).  
If the embedded Ising problem is formulated wrongly, it either might yield optimal solutions which are suboptimal for the original problem or, 
even worse, the solutions might not even be `de-embeddable', which means that they have no clear correspondence to any original solution.
An example for the latter is a chain of qubits where we get the solution values -1 for one half and +1 for the other. 
This can be addressed by applying a `\emph{strong coupling}' to vertices that belong to the embedding of a single original vertex 
to enforce that they behave collectively during the annealing process. 
We say they shall be \emph{synchronized}.  
This can be achieved by large absolute strengths on the edges between the vertices. 
But what is `strong enough'? 
V.~Choi has called this non-trivial problem of finding suitable parameters for the provable equivalence of the original and the embedded Ising problem 
the \emph{parameter setting} problem~\cite{choi2008minor}.   

Unfortunately, in practice, we need to take further restrictions on the parameters into account.
First of all, they can only be chosen within a certain interval,  
where the specific boundary values might vary between the different architectures or even devices.
At first sight, this might not appear to be problematic: 
We can simply scale the Ising problem by multiplying by a constant factor. 
This, however, decreases the absolute difference between the problem parameters,
while the most critical restriction of D-Wave's annealing machines is their parameter precision. 
Due to the transmission over the analog control circuits, the problem-defining parameters experience different perturbations~\cite{king2014algorithm}.  
This means that the actually solved problem differs slightly from the one specified by the user. 
Thus, problems which shall be solved with these machines need to be chosen carefully to yield some kind of `robustness' in the parameter precision.
 
Although the programming interface allows to insert arbitrary float values within given ranges, 
the machine can actually realize only a limited discrete parameter range.  
In~\cite{stollenwerk2019quantum}, a precision of about~$\frac{1}{30}$ was estimated for the specific annealer used in the experiments, 
which in turn means integer values between $-30$ and $30$ for a scaled problem. 
For problems with a higher precision, respectively larger integer parameters, the success probability is drastically reduced
because the annealing machine is not capable of resolving the parameters.  
In more recently released machines, the precision has probably been improved.    
However, the specific values and boundaries are not precisely known and can only be estimated through further experiments. 

For the users, the programming of such annealing machines is only worth the effort
if the machine can find the optimal solution to the provided problem in a certain number of runs, 
that is, if an acceptable \emph{success probability} can be achieved.
With the concrete restrictions on the internally implemented parameters not being specified exactly, 
we can therefore merely formulate some objectives aiming to improve the parameter distribution of the encoded Ising problem as much as possible, 
and thereby hopefully also the success probability.   
Because two parameters might appear too close to each other for the machine in presence of a very large parameter,
a first step is therefore to keep the largest appearing parameter as small as possible (without scaling).
This already concerns the encoding of an arbitrary combinatorial problem as a general Ising problem 
but becomes particularly important when the Ising problems shall be embedded:  
Such large values usually appear with the strong coupling of the embedded vertices. 
Therefore, the coupling strength cannot be chosen arbitrarily large.
 
Consequently, we do not only need to find a feasible parameter setting, 
ensuring the synchronization of the embedded vertices, %
but it also needs to be optimal in the sense that the coupling strength is as small as possible to conform with the precision of the machine. 
Only if this problem is solved, we can provide suitable embedded Ising problems and thus run meaningful experiments with the quantum annealers.
Furthermore, this only enables to analyze the actual performance of the machines
because miss-specified problems are not mixed up with the physical effects anymore,  
both suppressing the success probability in different ways.

\subsection{Related Work}

The baseline for all the work around minor embedding and the corresponding parameter setting was developed by V. Choi.
In \cite{choi2008minor}, a first upper bound on the strengths on the coupling edges depending on the original parameters is given,
achieved by providing an explicit non-uniform weighting of the vertices in the hardware graph. 
However, in practice, these bounds seem to be too weak and the large strengths they introduce suppress the success probability due to the necessary scaling factor.
Besides that, the explicit parameter setting problem is studied less intensively than the embedding problem, in particular analytically, 
although the limitations are quite well examined and understood and the choice of the strengths in the single vertex embeddings was recognized early to be decisive for the success probability of the D-Wave machine~\cite{king2014algorithm}.

By now, there is a common understanding in the quantum annealing community that the \emph{coupling strength}, 
the single strength value that is in most cases simply applied to all coupling edges, 
needs to be larger than the largest absolute parameter of the original Ising problem, 
but should not be orders of magnitudes larger to not trigger the precision issues of the annealer. 
Usually, a factor of 2 is applied, as for instance is described in~\cite{raymond2020improving}. 
At the same time, the weights are in general distributed uniformly over the vertices.  

Another method used in practice is determining the scaling factor empirically, see e.g.~\cite{venturelli2015quantum}. 
This means that several instances of the same original problem are transferred into Ising problems, usually yielding a common structure.
By successively solving the problems with certain parameters and checking the feasibility of the found solutions afterwards,
a specific bound or a bounding function in the input parameters is estimated and assumed to hold also for all other instances of the same problem.
In~\cite{pudenz2016parameter} different coupling strength scaling is tested with several strategies to chose the weights, but none of them shows a significant advantage over the other. 
In any case, such scanning does not provide any provable equivalence of the embedded Ising problem but can only give some guidelines.

In the package~\texttt{dwave-system}, D-Wave's programming interface offers a method to set the coupling strength called `uniform torque compensation'~\cite{dwave2022chainstrength}, 
which is most likely based on~\cite{raymond2020improving}. 
In the given formulation, it only applies for chains, which means if the embedding of a single vertex induces a path in the hardware graph. 
The method is derived from the idea that a `torque' on the central edge of the chain, caused by the supposedly random influence of the neighboring chains,
needs to be compensated by setting the weights and strengths accordingly. 
Although the results of the empirical study for certain random instances in~\cite{raymond2020improving} are promising, 
an analytical study of the equivalence of the thus obtained solutions is missing,
which is why this method can also only be considered as a heuristic approach to obtain the coupling strength.

The more recent publication~\cite{fang2020minimizing} is the first and only one after Choi's, to our best knowledge,  
that provides an analytical investigation of the general parameter setting problem.
Based on arbitrary, but given and fixed weights, the authors derive bounds on the coupling strength 
and show that their bounds are stronger than those of Choi and tight for some special cases. 

\subsection{Contribution}

In this work, we focus on the specific programming restrictions of the annealing machines, but, apart from that, 
we consider the annealers as a black box without questioning their ability to actually solve the programmed problems.
We aim to clearly divide the transformation steps of the problems towards the machine and close the loop to the embedded Ising problem 
before the annealers even are involved. %
Therefore, we answer a purely mathematical question here, that is interesting for itself, and thereby improve the application of quantum annealers. 

We provide a mathematical description of an embedded Ising problem %
that holds a provable equivalence to the original Ising problem,
which means both problems yield equivalent solutions. 
This includes embeddings that contain arbitrary embedded subgraphs rather than only chains as in previous approaches. 
By concentrating on synchronized solutions, we formulate general sufficient requirements. 
The observation of single vertices with their corresponding embeddings and certain assumptions on the thus extracted instances 
allow us to formulate specific constraints on the coupling strengths. 

Indeed the bounds of \cite{fang2020minimizing} look similar to the cut constraints which we derive in \autoref{sec:simple}. 
However, there is a major difference: Our constraints do not include the absolute values, 
which can be a decisive factor regarding the complexity of the problem. 
Our top-down approach, with a detailed deduction of our bounds, allows to clearly indicate why we can omit the absolute values.
With this we also prove the sufficiency of more general conditions on an embedded Ising problem.
We further state where we `lose the necessity' but can only derive the sufficiency of our requirements. 
Therefore, instances for which the bounds are tight can be identified more easily. 

By the choice of specific objective functions and the inclusion of a variable setting of the weights together with an additional gap parameter, 
we take a significant step further and extend the problem to a linear optimization problem yielding the optimal coupling strength. 
As such, we provide the first approach of analyzing the parameter setting problem in terms of mathematical optimization. 
We show by the reduction of the number of constraints that it is a problem which, in contrast to the embedding problem, can be solved easily, that is, in polynomial time
if the embedded vertices induce trees~\cite{lobe2022diss}.

\subsection{Structure}

First, we introduce the basic terms and concepts in \autoref{sec:basics}. 
After recapturing the main graph-theoretical terms used in this article in \autoref{sec:notation}, 
we provide an accurate background for the two main concepts in quantum annealing, the Ising problem and the graph embedding, in Sections~\ref{sec:ising} and~\ref{sec:emb}, respectively. 
Combining both concepts, we can establish the embedded Ising problem in \autoref{sec:embising}.
The strategy of synchronized variables is presented in \autoref{sec:synch}.  

In \autoref{sec:extract}, we break down the full embedded Ising problem into smaller problems, which can be solved individually: 
By extracting the part concerning a single vertex, we derive sufficient requirements on the parameters concerning this vertex in \autoref{sec:single}.
We formulate and simplify the corresponding optimization problem in \autoref{sec:instance} and \autoref{sec:simple}.
The resulting problem is summarized in \autoref{sec:probs}. 

The problem is then analyzed in \autoref{sec:analysis}. 
We establish a simplified description of the polyhedron over which it is defined in \autoref{sec:poly}.
By reducing the number of constraints significantly due to redundancy in \autoref{sec:connected}, we can derive the polynomial-time solvability for trees. 
Finally, we conclude our results in \autoref{sec:conclusion}.

\section{Basic Terms}\label{sec:basics}

\subsection{General Notation}\label{sec:notation}

First, we introduce some general notations used throughout this work.
For the basic graph definitions, we generally follow the standard literature in graph theory and optimization, see e.g.~\cite{diestel2017graph} or~\cite{korte2008combinatorial}, 
and briefly recapture the main notations here:
With $G = (V, E)$ we always refer to a simple undirected finite graph with the finite set of vertices $V$ and the set of edges $E \subseteq \{\{v, w\} : v, w \in  V\}$.
Given a graph $G$, $V(G)$ and $E(G)$ provide the vertex and the edge set, respectively, if those are not named specifically. 
While a subgraph of $G$ is formed by arbitrary subsets of edges and vertices of $G$,   
$G[S]$ refers to the vertex-induced subgraph of graph~$G$ for some vertex set $S \subset V(G)$, where we have $V(G[S]) = S$ and $E(G[S]) = \{\{v, w\} \in E(G) : v, w \in S\}$.  
For shortness, we abbreviate an edge $\{v, w\}$ with the commutative product $vw$. 
We denote the neighbors of a vertex $v$ in the graph $G$ with
\begin{equation}
    N(v) \vcentcolon = \{w \in V(G) : vw \in E(G)\}.
\end{equation} 
The incident edges are
\begin{align}
    \delta(S) &\vcentcolon = \{ vw \in E(G) : v \in S, w \in V(G)\setminus S\}, \\
    \delta(S, T) &\vcentcolon = \{ vw \in E(G) : v \in S, w \in T\} = \delta(S) \cap \delta(T),
\end{align}
where we use $\delta(v)$ to abbreviate $\delta(\{v\})$.

For indexed parameters or variables $x \in X^I$ with the index set $I$ and the value set $X$, 
we use $x_J = (x_i)_{i \in J}$  for a subset $J \subseteq I$ of the indices to refer to a subset of these parameters or variables, respectively, the corresponding vector. 
In turn, we `apply' $J$ by 
\begin{equation}
    x(J) = \sum_{i \in J} x_i. 
\end{equation}
We denote the vector containing only 1's or 0's by $\One$ and $\Zero$, respectively. 
For both, we add the subscript for the corresponding index set wherever necessary.   
If a set~$S$ is the disjoint union of two sets $S_1$ and $S_2$, that means $S_1 \cup S_2 = S$ and $S_1 \cap S_2 = \emptyset$, we use $S = S_1 \cupdot S_2$. 
With $2^X$ we denote the set of all subsets of a set~$X$.

\subsection{Ising Problem}\label{sec:ising}
 
In the quantum annealing processor, the magnetism of the superconducting loops and their couplings can be adjusted with user-defined input parameters.
This means we can encode different quadratic functions. %
The term `Ising model' also refers to these objective functions
because they are closely related to the formulation of the physical model~\cite{choi2008minor}.
We use throughout this work: 

\begin{definition}\label{def:ising}
    An \emph{Ising model} over a graph $G$ with \emph{weights} $\pmb{W} \in \R^{V(G)}$ and \emph{strengths} ${\pmb{S} \in \R^{E(G)}_{\neq 0}}$ is a function $\pmb{I_{W, S}} :\{-1, 1\}^{V(G)} \to \R$ with 
    \begin{equation}
        I_{W, S}(s) \vcentcolon = \sum_{v \in V(G)} W_v s_v + \sum_{vw \in E(G)} S_{vw} s_v s_w.
    \end{equation} 
    We call $G$ the \emph{interaction graph} of the Ising model. 
\end{definition}

Usually, we keep the interaction graph fixed. 
To be able to differ between two Ising models for the same graph, we use the symbol $I_{W, S}$ with the corresponding weights and strengths in the subscript.
In case those are clear from the context, we drop the subscript.
Using this definition, we can formulate a general version of the optimization problem the quantum annealing machine can process: 

\begin{fprob*}[\textsc{Ising Problem}]\label{prob:ising}
    Given a graph $G$, $W \in \R^{V(G)}$ and $S \in \R^{E(G)}$, find $s$ that solves
    \begin{equation}
        \min_{s \in \{-1,1\}^{V(G)}} I_{W, S}(s).
    \end{equation}
\end{fprob*}

D-Wave's quantum annealer can indeed only implement float values with $W \in [-m, m]^{V(G)}$ and $S \in [-n, n]^{E(G)}$ for specific $m, n \in \N$. 
For instance, for the current Chimera architecture, we have $m=2$ and $n=1$. 
However, due to possible scaling, this is not a hard restriction. 
A value which provides more insight in the coefficient distribution is the maximal absolute coefficient
\begin{equation}\label{eq:cmax}
    \pmb{C_{\max}} \vcentcolon = \max\big\{\|W\|_\infty,\, \|S\|_\infty\big\} = \max\left\{ \max_{v \in V(G)} |W_v|,\, \max_{vw \in E(G)} |S_{vw}|, \right\}, 
\end{equation}
in particular when compared with its counterpart, the minimal absolute coefficient being unequal to zero, 
or the minimal difference between two absolute coefficients
\begin{equation}
    \min\big\{ |x - y| : x \neq y \in \{0\} \cup \{W_v : v \in V(G)\} \cup \{S_{vw} : vw \in E(G)\}\big\}.
\end{equation}

If we further restrict the weights and strengths to $\Z$ according to the differentiation considerations, 
which means on the integer range $\{-m, -m + 1, ..., m\}$, respectively, $\{-n, -n+1, ..., n\}$, the latter becomes 1 after scaling.
Thus, the maximal absolute coefficient $C_{\max}$ is a decisive value to estimate 
whether the problem meets the parameter restrictions and is thus suitable to be solved with the annealer. 
According to~\cite{stollenwerk2019quantum}, we need at least $C_{\max} \leq 30$ to achieve an acceptable success probability.   

The decision problem corresponding to \refProb{ising} is known to be NP-complete~\cite{barahona1982computational}. 
This means a variety of problems can be mapped to it in polynomial time~\cite{lucas2014ising}. 
In particular, it is closely related to the \textsc{Quadratic Unconstrained Binary Optimization Problem}~(QUBO), 
more commonly known and well studied in combinatorial optimization. 
See, for example, \cite{kochenberger2014unconstrained} for more details.
 
There are preprocessing methods for directly manipulating the Ising model. 
One of them is applicable if the weight of a vertex exceeds the influence of the strengths of the incident edges.
We recall the well-known result here because it implies the exclusion of a certain weight-strengths constellation in the following investigations. 
Although it is already used in \cite{choi2008minor}, it is not formally proven there. 
Therefore, we also add the proof for completeness.  

\begin{lemma}\label{lem:preprocess}
    For an Ising model $I_{W, S} :\{-1, 1\}^{V(G)} \to \R$ over a graph $G$ with $W \in \R^{V(G)}$ and $S \in \R^{E(G)}$, 
    if we have 
    \begin{equation}
        |W_v| > \sum_{n \in N(v)} |S_{vn}| 
    \end{equation}
    for some vertex $v \in V(G)$, every optimal solution 
    \begin{equation}
        s^* \in \argmin_{s \in \{-1,1\}^{V(G)}} I_{W, S}(s)
    \end{equation}
    fulfils $s^*_v = - \sign(W_v)$. 
\end{lemma}
\begin{proof}
    We extract the part of $I_{W, S}$ containing $s^*_v$ with 
    \begin{equation}
        I_{W, S}(s^*) = \sum_{w \in V(G) \setminus \{v\}} W_w s^*_w + \sum_{wu \in E(G) \setminus \delta(v)} S_{wu} s^*_w s^*_u
                            + \underbrace{W_v s_v^* + \sum_{n \in N(v)} S_{vn} s_v^* s^*_n}_{=\vcentcolon{}I^v(s^*_v)} ,
    \end{equation}
    where we keep the other $s$-variables apart from $s^*_v$ fixed.
    With the condition for vertex $v$, we have 
    \begin{equation}
        |W_v| > \sum_{n \in N(v)} t_n S_{vn} \quad \forall t \in \{-1, 1\}^{N(v)}
    \end{equation} 
    and therefore can observe that
    \begin{equation}
    \begin{aligned}
        I^v(\sign(W_v)) &= |W_v| + \sum_{n \in N(v)} S_{vn} \big(\sign(W_v) s^*_n\big) \\ %
            &> 0 \\
            &> -|W_v| + \sum_{n \in N(v)} S_{vn} \big(- \sign(W_v) s^*_n\big) = I^v(-\sign(W_v)).
    \end{aligned}
    \end{equation}
    This shows that the contribution of $s^*_v = \sign(W_v)$ is always larger than the negated choice
    independently of the assignment of the other $s$-variables.  
\end{proof}

Remark: It is also easy to see that, if the equality holds in the above condition for $v$, 
the optimal solution does not necessarily hold the value $- \sign(W_v)$ for $s^*_v$. 
Still, this only happens if the last inequality in the proof collapses to an equality. 
Therefore, both choices yield the same optimal value and we can nevertheless choose to set $s^*_v = - \sign(W_v)$ in advance. 

Based on this result, we could remove certain variables from our Ising problem in advance. 
Therefore, we assume in the following that our given Ising model is not preprocessable according to the lemma anymore, that is, we have 
\begin{equation}\label{eq:unpreprocessable}
    |W_v| < \sum_{n \in N(v)} |S_{vn}|
\end{equation}
for all vertices $v \in V(G)$. \newpage

However, when solving problems with D-Wave's annealing machines, we cannot choose the interaction graph $G$ arbitrarily. 
It needs to correspond to the currently operating hardware graph. 
Only if $G$ is a subgraph of the hardware graph, we can directly solve \refProb{ising} with the D-Wave annealer (with some probability)
by setting surplus parameters to 0.

\subsection{Graph Embedding}\label{sec:emb}

D-Wave's quantum annealers do not realize fully connected graphs, 
which would allow for solving Ising models with arbitrary interaction graphs with the same or a smaller number of vertices.
They rather provide specific hardware graphs, representing the connectivity of the overlapping superconducting loops which form the qubits. 
For currently operating hardware, those are the \emph{Chimera} and \emph{Pegasus graphs}~\cite{dwave2022docs}.

If the investigated application is not explicitly customized to fit those graphs, 
the interaction graph of the corresponding Ising model does in most cases not have any relation to them. 
Thus, to be able to calculate on such annealing machines, we always have to deal with the discrepancy between the problem graphs and the realized hardware graphs:
We require what is known as an \emph{embedding}.   
That means several hardware vertices are combined to form a logical vertex to simulate an arbitrary problem connectivity. 
As we base the following work on it, we repeat and slightly extend the definition of \cite{lobe2021minor} here for completeness: 

\begin{definition}\label{def:embedding}
    For two graphs $G$ and $H$, an \emph{embedding} of $G$ in $H$ is a map $\pmb{\varphi}: V(G) \to 2^{V(H)}$ 
    fulfilling the following properties, where we use $\pmb{\varphi_v} \vcentcolon= \varphi(v)$ for $v \in V(G)$ for shortness: 
    \begin{enumerate}[a)]
        \item all $\varphi_v$ for $v \in V(G)$ induce disjoint connected subgraphs in $H$, more precisely {\setlength\parskip{0pt}
        \begin{itemize}[nosep]
            \item we have $\varphi_v \cap \varphi_w = \emptyset$ for all $v \neq w \in V(G)$ and
            \item $H[\varphi_v]$ is connected for all $v \in V(G)$,
        \end{itemize}}
        \item for all edges $vw \in E(G)$, there exists at least one edge in $H$ connecting the sets $\varphi_v$ and~$\varphi_w$, 
              which means we have $\delta(\varphi_v, \varphi_w) \neq \emptyset$.
    \end{enumerate}  
    We call $G$ \emph{embeddable} into $H$ if such an embedding function for $G$ and $H$ exists.
\end{definition}

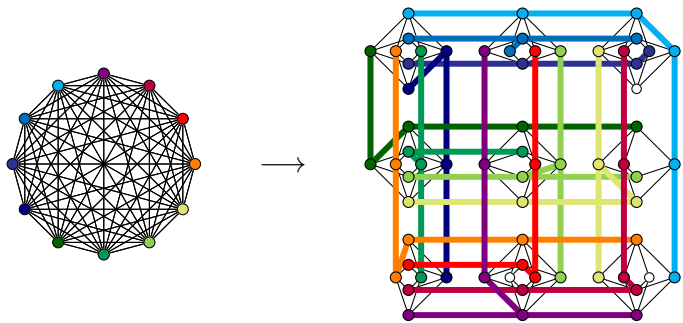
\begin{figure}[b!]
    \centering
    \def\scale{0.2}
    $\vcenter{\hbox{\begin{externalize}
\begin{tikzpicture}[scale=\scale]

    \foreach \i in {1, ..., 12} {
            \setcolor{\i}
            \node[nodec, fill=currentcolor, scale=1.25] (\i) at ($(2,-2)+(90+\i*360/12:6)$) {};
    }

    \foreach \v in {1,...,12}{
    \foreach \w in {1,...,12}{
        \draw (\v) -- (\w);
    }}

\end{tikzpicture}
\end{externalize}}}$ $\quad \longrightarrow \quad$
    \def\scale{0.25} 
    $\vcenter{\hbox{\begin{externalize}
\begin{tikzpicture}[scale=\scale]

    \setboolean{drawBrokenQubits}{false}
    \setBrokenQubits{{(1-2-v4), (3-1-h1), (2-2-h2), (2-3-v2), (2-3-h3), (3-3-v2)}}
    \drawBrokenChimera

    \drawCrossFromToCC{1}{3}{1}{4}{1}{3}{1}{3}
    \drawCrossFromToCC{1}{2}{2}{2}{1}{3}{1}{1}
    \drawCrossFromToCC{1}{3}{3}{3}{1}{3}{1}{1}
    \drawCrossFromToCC{1}{1}{4}{4}{1}{1}{1}{3}

    \drawCrossFromToCC{2}{1}{1}{1}{1}{3}{1}{2}
    \drawCrossFromToCC{2}{1}{2}{3}{1}{2}{1}{3}
    \drawCrossFromToCC{2}{2}{3}{4}{1}{3}{1}{3}
    \drawCrossFromToCC{2}{3}{4}{1}{1}{3}{1}{3}

    \drawCrossFromToCC{3}{1}{1}{2}{1}{3}{1}{3}
    \drawCrossFromToCC{3}{2}{2}{3}{1}{2}{1}{3}
    \drawCrossFromToCC{3}{3}{3}{2}{1}{3}{1}{3}
    \drawCrossFromToCC{3}{2}{4}{1}{1}{3}{1}{3}

\end{tikzpicture}
\end{externalize}}}$
    \caption{Exemplary complete graph embedding in a broken Chimera graph \cite{lobe2021embedding}}
    \label{fig:embedding}
\end{figure}

An example of such an embedding is shown in \autoref{fig:embedding}. 
The concept of embeddings is closely related to graph minors, which is why they are also called \emph{minor embeddings}~\cite{choi2011minor}.
General graph minors have been intensively studied even before quantum annealing became a hot topic
and the basis is formed by Robertson and Seymore, see e.g.~\cite{robertson1995graph}.
In the quantum annealing context, $H$ refers to the hardware graph, such as a Chimera graph, 
while $G$ is the problem graph derived from the specific application and its concrete Ising formulation, which can therefore be fully arbitrary.
In the following work, we consider the embedding to be given. 

\subsection{Embedded Ising Problem}\label{sec:embising}
   
The Ising model as given in \autoref{def:ising} is defined over arbitrary graphs, thus also over the possible hardware graphs. 
As explained before, typical applications however need an embedding.  
Therefore, we introduce an extended definition of the Ising model in this section to combine both concepts. 
For this we first extend the embedding notation of \autoref{def:embedding} by the following graph structures: 

\begin{definition}\label{def:embgraph}
    For two graphs $G$ and $H$ and an embedding $\varphi: V(G) \to 2^{V(H)}$ of $G$ in $H$, let the \emph{embedded graph}, 
    the subgraph of $H$ resulting from the embedding, be 
    \begin{align}
        \pmb{H_\varphi} &\vcentcolon = H\left[\bigcup_{v \in V(G)} \varphi_v\right] = \left(\bigcup_{v \in V(G)} \varphi_v, E_\varphi \cupdot E_\delta\right)
    \intertext{with}
        \pmb{E_\varphi} &\vcentcolon = \bigcup_{v \in V(G)} E(H[\varphi_v]), \\
        \pmb{E_\delta} &\vcentcolon = \bigcup_{vw \in E(G)} \pmb{\delta_{vw}} \vcentcolon = \bigcup_{vw \in E(G)} \delta(\varphi_v, \varphi_w),
    \end{align}
    denoting the \emph{intra-connecting} and the \emph{inter-connecting edges}, respectively.
\end{definition}

Using the embedding objects of \autoref{def:embgraph}, we can now formulate an Ising model over the given embedded graph. 
The following concepts are mainly well known in the quantum annealing community, see e.g.~\cite{choi2008minor} and \cite{raymond2020improving}, but we want to bring them into a more formal format here. 
    
\begin{definition}
    An \emph{embedded Ising model} for two graphs $G$ and $H$ and an embedding ${\varphi: V(G) \to 2^{V(H)}}$ of $G$ in $H$ is an Ising model over $H_\varphi$, 
    where we have $\pmb{\closure{I}_{\closure{W}, \closure{S}}} :\{-1, 1\}^{V(H_\varphi)} \to \R$ with 
    \begin{equation}
    \begin{aligned}
        \closure{I}_{\closure{W}, \closure{S}}(s) &\vcentcolon = \sum_{v \in V(G)} \left(\sum_{q \in \varphi_v} \closure{W}_q s_q + \!\sum_{pq \in E(H[\varphi_v])} \!\closure{S}_{pq} s_p s_q\right) 
                         + \sum_{vw \in E(G)} \sum_{pq \in \delta_{vw}}  \closure{S}_{pq} s_p s_q \\
                       &\phantom{\vcentcolon} = \sum_{q \in V(H_\varphi)} \closure{W}_q s_q + \sum_{pq \in E_\varphi \cup E_\delta} \closure{S}_{pq} s_p s_q   
    \end{aligned}  
    \end{equation}
    for the weights $\pmb{\closure{W}} \in \R^{V(H_\varphi)}$ and the strengths $\pmb{\closure{S}} \in \R^{E(H_\varphi)}$.
\end{definition}   

In this case, we call the corresponding \refProb[]{ising} of finding an $s$ that solves
\begin{equation}
    \min_{s \in \{-1,1\}^{V(H_\varphi)}} \closure{I}_{\closure{W}, \closure{S}}(s)
\end{equation}
with the above embedded Ising model $\closure{I}_{\closure{W}, \closure{S}}$ the~\textsc{Embedded Ising Problem}\namedlabel{prob:embising}{\textsc{Embedded Ising Problem}}. 
\addcontentsline{loe}{prob*}{\textsc{Embedded Ising Problem}}

With this formulation and $H = C$ for $C$ being a currently operating broken Chimera graph of \mbox{D-Wave}, 
we could solve the corresponding \refProbNamed[]{embising} with the D-Wave annealer (with some probability). 
However, given an arbitrary Ising model whose underlying connectivity graph requires an embedding, 
we need to find a suitable corresponding embedded Ising model. 
This requires to choose the weights and strengths in a certain way
such that an optimal solution of the new \refProb[]{ising} corresponds to an optimal solution of the original one in the end. 

In particular, as we usually do not only want to know the optimal value but also the optimal solution itself, 
we need a recipe how to get from an embedded to an original solution. 
We therefore need a `de-embedding' function that can be computed easily, which means in polynomial time. 
This is more formally stated by the following definition, where we drop the weights and the strengths in the subscript for simplicity. 

\begin{definition}\label{def:equivalent}
    An \emph{equivalent embedded Ising model} $\closure{I} :\{-1, 1\}^{V(H_\varphi)} \to \R$ to a given Ising model $I :\{-1, 1\}^{V(G)} \to \R$ 
    for two graphs $G$ and $H$ and an embedding ${\varphi: V(G) \to 2^{V(H)}}$ of $G$ in $H$ fulfils the following properties: 
    \begin{itemize}[a)]
        \item The corresponding Ising problems are equivalent in the sense that we have 
            \begin{equation}
                \min_{s \in \{-1,1\}^{V(H_\varphi)}} \closure{I}(s) + c = \min_{t \in \{-1,1\}^{V(G)}} I(t)
            \end{equation}
            for a known constant $c \in \R$ and
        \item there exists a mapping from an optimal solution $s^* \in \{-1,1\}^{V(H_\varphi)}$ of \refProbNamed{embising} to 
            an optimal solution $t^* \in \{-1,1\}^{V(G)}$ of the (unembedded) \refProb[]{ising} which can be computed in polynomial time. 
    \end{itemize}
    
\end{definition}
 
This would have been sufficient to use the quantum annealing machines if the underlying physical system had ideally realized the corresponding physical model.    
However, this is impossible in the real world and the machines thus only work heuristically providing solutions only with some unknown probability. 
In general, it remains unclear whether we have found the optimal solution, a sub-optimal solution or no solution at all.
Thus, the user does not only need to have access to the mentioned mapping of the optimal solutions but rather needs more information to deal with the results of the machine. 

In practice, we need an extended version of the above definition to overcome this issue: 
For each solution provided by the annealer, not only optimal ones, we want to know whether we can de-embed it to an original solution and 
if we can, we also want to know how to do it.
We define: 
 
\begin{definition}
    An equivalent embedded Ising model $\closure{I} :\{-1, 1\}^{V(H_\varphi)} \to \R$ to a given Ising model $I :\{-1, 1\}^{V(G)} \to \R$ 
    for two graphs $G$ and $H$ and an embedding ${\varphi: V(G) \to 2^{V(H)}}$ of $G$ in $H$ is called \emph{de-embeddable} 
    if we have two functions  
    \begin{align}
        \pmb{\psi} : \{-1,1\}^{V(H_\varphi)} &\to \{0, 1\}
    \shortintertext{and}
        \pmb{\tau} : \big\{s \in \{-1,1\}^{V(H_\varphi)} : \psi(s) = 1\big\} &\to \{-1,1\}^{V(G)}
    \end{align}
    which can both be computed in polynomial time.
    While
    \begin{equation}
        \psi(s) = \begin{cases}
            1 &\text{if $s$ is de-embeddable}, \\
            0 &\text{otherwise}
        \end{cases}    
    \end{equation}
    tells whether we can compute an original solution to the embedded one, the function $\tau$ provides the corresponding de-embedded solution, 
    where we have 
    \begin{equation}
        \closure{I}(s) + c = I(\tau(s)) \qquad\forall s \in \{-1,1\}^{V(H_\varphi)} \text{ with } \psi(s) = 1
    \end{equation}
    for the constant $c \in \R$.  
    
    We call
    \begin{equation}
        \psi^{-1}(1) = \big\{s \in \{-1,1\}^{V(H_\varphi)} : \psi(s) = 1\big\}
    \end{equation} 
    the set of \emph{de-embeddable solutions}.   
\end{definition}

Thus, for 
\begin{equation}
   s^* \in \argmin_{s \in \{-1,1\}^{V(H_\varphi)}} \closure{I}(s),
\end{equation}
we have by \autoref{def:equivalent}
\begin{equation}
   \tau(s^*) \in \argmin_{t \in \{-1,1\}^{V(G)}} I(t).
\end{equation}   
The most useful in practice would be if all original solutions had a corresponding embedded counterpart,
which means if $\tau$ is surjective.
This in turn would mean we have $\psi^{-1}(1) \cong \{-1,1\}^{V(G)}$ and at least $2^{|V(G)|}$ solutions that are de-embeddable. 

To find such functions, we need to decide at some point what structure the embedded solutions should follow.
Although different options might be possible due to the large number of adjustable parameters, 
the most straightforward way is to restrict the considerations to solutions where all variables corresponding to the embedding of a single original vertex hold the same value. 
This principle called \emph{synchronization} is explained in the following section in more detail.

\subsection{Variable Synchronization}\label{sec:synch}

The main aspect of the equivalence of the given and \refProbNamed{embising} is the retrieval of the original solution from the embedded one.
For this we need to be able to `de-embed' the embedded solution. 
This in turn requires this solution to hold a certain structure.
By enforcing the \emph{synchronization} of all variables in the embedded Ising model that correspond to a single original variable,
which means that all those variables should hold the same value, 
we have a simple criterion on the solutions of \refProbNamed{embising}. 
This idea was already introduced in \cite{choi2008minor} and means more formally

\begin{definition}\label{def:synch}
    A solution of \refProbNamed{embising} $s \in \{-1, 1\}^{V(H_\varphi)}$ is called a \emph{synchronized solution} 
    with respect to an embedding ${\varphi: V(G) \to 2^{V(H)}}$ of $G$ in $H$ for two graphs $G$ and $H$
    if we have 
    \begin{equation}
        s_q = t_v \in \{-1, 1\} ~ \forall q\in \varphi_v ~\forall v \in V(G). %
    \end{equation}  
\end{definition}

For such a synchronized solution, we can easily provide the functions required for the de-em\-bed\-ding with
\begin{equation}
    \psi(s) = \begin{cases}
        1 &\text{if } s_q = s_p ~\forall p, q \in \varphi_v ~\forall v \in V(G), \\
        0 &\text{otherwise}
    \end{cases}
\end{equation}
and 
\begin{equation}
    \tau(s) = s_{X}
\end{equation}
for some vertex set $X \subseteq V(H_\varphi)$ with $|X \cap \varphi_v| = 1$ for all $v \in V(G)$.  
The vertex set $X$ just serves as a placeholder, as we can simply choose a random vertex from $\varphi_v$ to obtain the value of its variable
because all of them hold the same value.  
It is easy to recognize that $\tau$ is surjective and both functions can be computed in polynomial time.  

In case the embedded variables do not hold a common value, it is unclear which value to assign to the corresponding original variable. 
In such cases, the common practice is to apply a post-processing on these unsynchronized solutions.
A popular example is the heuristic of \emph{majority voting}, where the original variable gets the value 
which appears in the majority of the assignments of the embedded variables \cite{king2014algorithm}. 
Those heuristics might be useful, when considering the non-optimal solutions provided by the D-Wave machine due to its physical `imperfectness'.
That means, for instance, if only a few variables are flipped in the found solution compared to the optimal solution due to single-qubit failures or read-out errors.  

However, if the embedded Ising model is ill-defined, which means that its optimal solution does not yield a clear correspondence to an original solution, 
those heuristics will not be able to extract the optimal original solution:
Switching the value of an embedded variable, to the one of the majority, also changes the contribution of some edges by their strength to the objective value, 
which in turn influences the neighboring vertices.
Thus, broken embeddings might have a global impact on the assignment of a large number of variables, 
which can usually not be `repaired locally'. 
Applying such methods in these cases will therefore in general not increase the probability of finding the optimal solution.  
On the other hand, we do not see a way how to construct an embedded Ising 
tailored to obtain the provable equivalence to the original Ising problem under such `majority solutions' 
due to the large number of possible distributions. 

Thus, how do we ensure that such an \emph{embedded Ising model based on synchronization}, which means it yields the given functions of \autoref{def:synch} as a de-embedding, 
is an equivalent embedded Ising model to our given one? 
Obviously, the weights and the strengths of the embedded Ising model depend on the original parameters. 
 
If the weights and the strengths fulfil
\begin{align}
    W_v = \sum_{\mathclap{q \in \varphi_v}} \closure{W}_q
\shortintertext{and }
    S_{vw} = \sum_{pq \in \delta_{vw}} \closure{S}_{pq},
\end{align}
respectively, we have for a synchronized solution $s$ as given in \autoref{def:synch}
\begin{equation}
\begin{aligned}
    \closure{I}(s) &= \sum_{v \in V(G)} \left(\sum_{q \in \varphi_v} \closure{W}_q s_q + \sum_{pq \in E(H[\varphi_v])} \closure{S}_{pq} s_p s_q\right) 
                     + \sum_{vw \in E(G)} \sum_{pq \in \delta_{vw}}  \closure{S}_{pq} s_p s_q \\
                   &= \sum_{v \in V(G)} \left(\sum_{q \in \varphi_v} \closure{W}_q t_v + \sum_{pq \in E(H[\varphi_v])} \closure{S}_{pq} t_v t_v\right) 
                     + \sum_{vw \in E(G)} \sum_{pq \in \delta_{vw}}  \closure{S}_{pq} t_v t_w \\
                   &= \sum_{v \in V(G)} t_v \ifjournalversion{\sum_{q \in \varphi_v} \closure{W}_q}{\left(\sum_{q \in \varphi_v} \closure{W}_q \right)} 
                     + \sum_{v \in V(G)} \sum_{pq \in E(H[\varphi_v])} \closure{S}_{pq}
                     + \sum_{vw \in E(G)} t_v t_w \ifjournalversion{\sum_{pq \in \delta_{vw}} \closure{S}_{pq}}{\left(\sum_{pq \in \delta_{vw}} \closure{S}_{pq} \right)}\\
                   &= \sum_{v \in V(G)} W_v t_v + \sum_{vw \in E(G)} S_{vw} t_v t_w + \sum_{pq \in E_\delta} \closure{S}_{pq}\\        
                   &= I(t) + \sum_{pq \in E_\delta} \closure{S}_{pq}.
\end{aligned}  \label{eq:synch}
\end{equation}
This means, for all such synchronized solutions, we have $\closure{I}(s) + c = I(t)$ with 
\begin{equation}
    c = - \sum_{pq \in E_\varphi} \closure{S}_{pq}. 
\end{equation}
Thus, the strengths $\closure{S}_{E_\varphi}$ only introduce an offset to the overall objective value for these solutions.   
Furthermore, we ensure that for an optimal solution
\begin{equation}
    t^* \in \argmin_{t \in \{-1,1\}^{V(G)}} I(t)
\end{equation}
we have $\closure{I}(s^*) + c = I(t^*)$ for $s^* = (t_v^* \One_{\varphi_v})_{v \in V(G)}$
and $s^*$ thus also is the minimum over all synchronized solutions, which means
\begin{equation}
    s^* \in \argmin \left\{ \closure{I}(s) : s \in \{-1,1\}^{V(H_\varphi)}, s_{\varphi_v} \in \{-\One, \One\} \forall v \in V\right\}. 
\end{equation}

However, for the given $s^*$, we do not necessarily have 
\begin{equation}
    s^* \in \argmin_{s \in \{-1,1\}^{V(H_\varphi)}} \closure{I}(s),
\end{equation}
which means it would also be the optimum over all solutions of \refProbNamed{embising}. 
There might be unsynchronized variable assignments yielding a lower objective value.
This is the case if the contribution of the inter-connecting edges does not suffice.
  
As it can be seen in \eqref{eq:synch}, if the variables $s_q$ and $s_p$ for $pq \in E(H[\varphi_v])$ are synchronized, 
their product reduces to 1 and the corresponding strength $\closure{S}_{pq}$ is added to the objective value.
In turn, if the variables are assigned to different values, the product is $-1$ and $\closure{S}_{pq}$ is subtracted. 
Due to the minimization, it is therefore preferable to set $\closure{S}_{pq}$ to a negative value. 
However, its contribution also needs to exceed the benefit of breaking the synchronization in the remaining part of the objective function.   

To ensure the synchronization, we could, in theory, set $\closure{S}_{pq} = - \infty$ for all $pq \in E_\varphi$ or at least to a very large negative value, 
e.g.\ exceeding the sum of the absolute values of all coefficients in the embedded Ising model. 
In this case, we could also choose $\closure{W}$ and $\closure{S}_{E_\delta}$ arbitrarily within the sum bounds. 
However, these large strength values cannot be realized in practice 
because the annealing machines have a limited parameter precision and height due to physical restrictions. 
Thus, how do we need to choose the parameters $\closure{S}_{E_\varphi}$ such that they suffice for the synchronization
and how does their choice influence possible choices for $\closure{W}$ and $\closure{S}_{E_\delta}$ and vice versa?

\section{Optimization Problem Extraction}\label{sec:extract}

    For calculations on the D-Wave machine, it is essential for the user that the encoded problem indeed represents the original problem the user wants to solve. 
    In this section, we extract and simplify the sufficient requirements on the parameters that need to be fulfilled 
    such that the resulting \refProbNamed[]{embising} provably holds equivalent solutions to those of the given problem, 
    based on the synchronization of all variables in the embedded problem corresponding to one variable of the original one.   
    By observing a single original vertex and adding an objective function aiming to minimize the absolute height of the parameters, 
    we can extract a specific optimization problem respecting the physical restrictions of the machine.  
    
    We assume the two graphs $\pmb{G}$ and $\pmb{H}$, the embedding $\pmb{\varphi}: V(G) \to 2^{V(H)}$ of $G$ in $H$ with the corresponding graph structures of \autoref{def:embgraph} 
    and an Ising model  $\pmb{I_{W, S}} :\{-1, 1\}^{V(G)} \to \R$ with the weights ${\pmb{W} \in \R^{V(G)}}$ and strengths $\pmb{S} \in \R^{E(G)}_{\neq 0}$ to be given and fixed in the following. 
    Given this data, how do we find an equivalent embedded Ising model $\closure{I}_{\closure{W}, \closure{S}} :\{-1, 1\}^{V(H_\varphi)} \to \R$  to $I_{W, S}$ 
    with weights $\closure{W} \in \R^{V(H_\varphi)}$ and strengths~$\closure{S} \in \R^{E(H_\varphi)}$? 
    Note that we drop the subscripts of the Ising models in most cases for simplicity.

\subsection{Single Vertex Evaluation}\label{sec:single}

To answer the question stated at the end of \autoref{sec:synch},  
we extract the part of the embedded Ising model that concerns a single original vertex $v \in V(G)$:
\begin{equation}
\begin{aligned}
    \closure{I}(s) ={}& \sum_{w \in V(G) \setminus \{v\}} \left(\sum_{q \in \varphi_w} \closure{W}_q s_q + \sum_{pq \in E(H[\varphi_w])} \closure{S}_{pq} s_p s_q\right) 
                      \ifjournalversion{\\ &}{} + \sum_{wu \in E(G) \setminus \delta(v)} \sum_{pq \in \delta_{wu}}  \closure{S}_{pq} s_p s_q \\
                      & + \underbrace{\sum_{q \in \varphi_v} \closure{W}_q s_q + \sum_{pq \in E(H[\varphi_v])} \closure{S}_{pq} s_p s_q
                        + \sum_{w \in N(v)} \sum_{pq \in \delta_{vw}}  \closure{S}_{pq} s_p s_q}_{= \vcentcolon \closure{I}\vphantom{I}^v(s)}. \\
\end{aligned}
\end{equation}
By this the remaining part $\closure{I}(s) - \closure{I}\vphantom{I}^v(s)$ %
does only depend on $s \in \{-1, 1\}^{V(H_\varphi) \setminus \varphi_v}$. 
By replacing $s \in \{-1, 1\}^{\varphi_v \cup N(\varphi_v)}$ with $(r, s) \in \{-1, 1\}^{\varphi_v} \times \{-1, 1\}^{N(\varphi_v)}$ in $\closure{I}\vphantom{I}^v(s)$
we get %
\begin{equation}
    \pmb{\closure{I}\vphantom{I}^v(r, s)} \vcentcolon = \sum_{q \in \varphi_v} \closure{W}_q r_q + \sum_{pq \in E(H[\varphi_v])} \closure{S}_{pq} r_p r_q
                        + \sum_{w \in N(v)} \sum_{pq \in \delta_{vw}}  \closure{S}_{pq} r_p s_q
\end{equation}
and can clearly indicate the different influencing parts.  
All variables corresponding to the embedding of vertex $v$, the $r$-variables, now only appear in $\closure{I}\vphantom{I}^v$, while the $s$-variables form the connection to the remaining part, 
thus appear in both $\closure{I} - \closure{I}\vphantom{I}^v$ %
and $\closure{I}\vphantom{I}^v$.  

In the following, we want to enforce the synchronization of the $r$'s independently of the influence `from the outside', which means for arbitrary $s$. 
Due to the minimization of the Ising models, this means that the minimum of the partial Ising problem should always be either $\One$ or $-\One$, more formally
\begin{equation}
    \argmin_{r \in \{-1,1\}^{\varphi_v}} \closure{I}\vphantom{I}^v(r, s) \subseteq \{-\One, \One\} \quad \forall s \in \{-1,1\}^{N(\varphi_v)}.
\end{equation}
In other words, we have
\begin{equation}
    \min_{r \in \{-1,1\}^{\varphi_v}} \closure{I}\vphantom{I}^v(r, s) = \min \big\{\closure{I}\vphantom{I}^v(-\One,s), \closure{I}\vphantom{I}^v(\One,s)\big\}
\end{equation}
but with
\begin{equation}
    \closure{I}\vphantom{I}^v(r, s) > \min \big\{\closure{I}\vphantom{I}^v(-\One, s), \closure{I}\vphantom{I}^v(\One, s) \big\} \quad\forall s \in \{-1,1\}^{N(\varphi_v)} ~\forall r \in \{-1,1\}^{\varphi_v} \setminus \{-\One, \One\}.
\end{equation}

Do these conditions applied to all vertices $v \in V(G)$ ensure that the embedded problem is provably equivalent to the original one?
We can indeed show their sufficiency:

\begin{lemma}\label{lem:single_ok}
    With
    \begin{equation}
        \closure{I}\vphantom{I}^v(r, s) > \min \big\{\closure{I}\vphantom{I}^v(-\One, s), \closure{I}\vphantom{I}^v(\One, s) \big\} \quad\forall s \in \{-1,1\}^{N(\varphi_v)} ~\forall r \in \{-1,1\}^{\varphi_v} \setminus \{-\One, \One\}
    \end{equation}
    for all $v \in V(G)$, 
    we have for all 
    \begin{equation}
        s^* \in \argmin_{s \in \{-1,1\}^{V(H_\varphi)}} \closure{I}(s)    
    \end{equation}
    that $s^* = (t_v^* \One_{\varphi_v})_{v \in V(G)}$ with $t^* \in \{-1, 1\}^V$.
\end{lemma}
\begin{proof}
    Assume there exists 
    \begin{equation}
        s^* \in \argmin_{s \in \{-1,1\}^{V(H_\varphi)}} \closure{I}(s)    
    \end{equation}
    with $s^*_{\varphi_v} \not\in \{-\One, \One\}$ for some vertex $v \in V(G)$. 
    Then we have 
    \begin{align}
        \closure{I}(s^*) 
            &= \closure{I}\vphantom{I}^v\Big(s^*_{\varphi_v}, s^*_{N(\varphi_v)}\Big) + \closure{I}\vphantom{I}^{V(G) \setminus \{v\}}\Big(s^*_{V(H_\varphi) \setminus \varphi_v}\Big) \\
            &> \min \Big\{\closure{I}\vphantom{I}^v\Big({-\One}, s^*_{N(\varphi_v)}\Big), \closure{I}\vphantom{I}^v\Big(\One, s^*_{N(\varphi_v)}\Big) \Big\} + \closure{I}\vphantom{I}^{V(G) \setminus \{v\}}\left(s^*_{V(H_\varphi) \setminus \varphi_v}\right) 
    \intertext{by the given conditions and we can further deduce}
            \closure{I}(s^*) &= \closure{I}\vphantom{I}^v\Big( r^*\One, s^*_{N(\varphi_v)}\Big) + \closure{I}\vphantom{I}^{V(G) \setminus \{v\}}\left(s^*_{V(H_\varphi) \setminus \varphi_v}\right) \\
            &= \closure{I} \big( \tilde{s}^*\big)
    \end{align}
    for $\tilde{s} \in \{-1,1\}^{V(H_\varphi)}$ with $\tilde{s}_{V(H_\varphi) \setminus \varphi_v} = s^*_{V(H_\varphi) \setminus \varphi_v}$ and $\tilde{s}_{\varphi_v} = r^* \One$ for 
    \begin{equation}
         r^* = \begin{cases}
            1 &\text{if } \closure{I}\vphantom{I}^v\Big({-\One}, s^*_{N(\varphi_v)}\Big) \geq \closure{I}\vphantom{I}^v\Big(\One, s^*_{N(\varphi_v)}\Big), \\
            -1 &\text{otherwise.} 
        \end{cases}
    \end{equation} 
    This contradicts to $s^*$ being an optimal solution. 
\end{proof}

With this result, we can now clearly formulate the requirements on an embedded Ising model: 

\begin{theorem}
    For two graphs $G$ and $H$, an embedding $\varphi  : V(G) \to 2^{V(H)}$ of $G$ in $H$ %
    and an Ising model  ${I_{W, S} :\{-1, 1\}^{V(G)} \to \R}$ with weights $W \in \R^{V(G)}$ and strengths $S \in \R^{E(G)}$, 
    the Ising model $\closure{I}_{\closure{W}, \closure{S}} :\{-1, 1\}^{V(H_\varphi)} \to \R$ with weights $\closure{W} \in \R^{V(H_\varphi)}$ and strengths $\closure{S} \in \R^{E(H_\varphi)}$ 
    forms an equivalent embedded Ising model to $I_{W, S}$ if we have 
    \begin{align}
        W_v &= \sum_{\mathclap{q \in \varphi_v}} \closure{W}_q &&\forall v \in V(G), \\
        S_{vw} &= \sum_{pq \in \delta_{vw}} \closure{S}_{pq} &&\forall vw \in E(G), \\
        \closure{I}\vphantom{I}^v_{\closure{W}, \closure{S}}(r, s) &> \min \left\{\closure{I}\vphantom{I}^v_{\closure{W}, \closure{S}}(-\One, s), \closure{I}\vphantom{I}^v_{\closure{W}, \closure{S}}(\One, s) \right\} 
            &&\forall s \in \{-1,1\}^{N(\varphi_v)} \\
            &&&\forall r \in \{-1,1\}^{\varphi_v} \setminus \{-\One, \One\} \\ 
            &&&\forall v \in V(G).
    \end{align}  
\end{theorem}
\begin{proof}
    The optimality is clear with the deductions from the beginning of this section and \autoref{lem:single_ok}. 
    Furthermore, from an optimal solution 
    \begin{equation}
        s^* \in \argmin_{s \in \{-1,1\}^{V(H_\varphi)}} \closure{I}(s),  
    \end{equation}
    we can easily get a solution of the original Ising problem with $t^*_v = s^*_q$ for an arbitrarily chosen $p \in \varphi_v$ for all $v \in V$ due to the enforced synchronization.  
\end{proof}

Note that this theorem only shows the sufficiency of our derived conditions.
However, it does not state anything about the necessity. 
In the constraints for a specific vertex $v$, we assume the \mbox{$s$-variables} to be fully arbitrary. 
If we took into account that some of them are not independent from each other as they belong to the embedding of a single neighbour of $v$, 
whose embedded vertices should equivalently be synchronized, we would possibly retrieve a stronger set of constraints. 
However, this introduces another level of complexity, which we keep for future research.

\subsection{Problem Instance Definition}\label{sec:instance}

In the following, we only concentrate on a single fixed vertex $\pmb{v} \in V(G)$. 
From the corresponding part of the embedded Ising model $\closure{I}\vphantom{I}^v : \{-1, 1\}^{\varphi_v} \times \{-1, 1\}^{N(\varphi_v)} \to \R$ 
and the constraints on the weights and strengths that concern $v$, 
we derive a specific optimization problem
that needs to be solved to obtain those parameter values
that ensure that the embedded vertices in $\varphi_v$ represent the original one $v$. 

\subsubsection{Input}

The part of the embedded graph $\closure{I}\vphantom{I}^v$ is working on is the \emph{embedded subgraph structure}
\begin{equation}
    H\left[\varphi_v \cupdot N(\varphi_v)\right] = (\varphi_v \cupdot N(\varphi_v), E(H[\varphi_v]) \cupdot \delta(\varphi_v) \cupdot E(H[N(\varphi_v)])),   
\end{equation} 
where we have
\begin{itemize}
    \item the connected inner graph $H[\varphi_v] = \vcentcolon \left(\vertices, \edges\right)$ with vertices $\pmb{\vertices} \vcentcolon = \varphi_v$ and edges $\pmb{\edges} \vcentcolon = E(H[\varphi_v])$ 
    \item the outer neighbors $\pmb{\outerneighbors} \vcentcolon = N(\varphi_v) \subseteq \bigcup \{p \in \varphi_w : w \in N(v)\}$, 
    \item the set of edges to the outer neighbors $\pmb{\outeredges} \vcentcolon = \delta(\varphi_v)$ and 
    \item the edges between the outer neighbors $E(H[N(\varphi_v)])$. 
\end{itemize}

\begin{figure}[b!]
    \centering
    \def\scale{0.2}
    $\vcenter{\hbox{\begin{externalize}
\begin{tikzpicture}[scale=\scale]

	\foreach \i in {1, ..., 12} {
			\setcolor{\i}
    		\node[nodec, fill=currentcolor, scale=1.25] (\i) at ($(2,-2)+(90+\i*360/12:6)$) {};
	}		

    \foreach \w in {1,...,12}{
        \draw (7) -- (\w);
    }

\end{tikzpicture}
\end{externalize}}}$ $\quad \longrightarrow \quad$
    \def\scale{0.25}
    $\vcenter{\hbox{\begin{externalize}
\begin{tikzpicture}[scale=\scale]

    \setboolean{drawBrokenQubits}{false}
    
    \foreach \r in {1, 3} {
        \foreach \c in {1, 3} {
            \foreach \d in {1, 2, 3, 4} {
                \forcsvlist{\listxadd\broken}{(\r-\c-h\d), (\r-\c-v\d)}
            }
        }
    }
    
    \setBrokenQubits{\broken}
    \addBrokenQubits{{(1-2-v4), (1-2-h1), (1-2-h2), (1-2-h3), 
                      (2-1-v1), (2-1-v2), (2-1-v4),
                      (2-2-h2), 
                      (2-3-v1), (2-3-v2), (2-3-v4), (2-3-h3),  
                      (3-2-h1), (3-2-h2), (3-2-h3)}}
    \drawBrokenChimera

    \setcolorCounter
    \node[nodec, fill=currentcolor] at (1-2-v1) {};
    \node[nodec, fill=currentcolor] at (2-3-h4) {};

    \setcolorCounter
    \node[nodec, fill=currentcolor] at (1-2-v2) {};
    
    \setcolorCounter
    \node[nodec, fill=currentcolor] at (1-2-v3) {};
    
    \setcolorCounter
    \node[nodec, fill=currentcolor] at (2-1-h4) {};

    \setcolorCounter
    \node[nodec, fill=currentcolor] at (2-1-h1) {};
    \node[nodec, fill=currentcolor] at (2-2-v1) {};

    \setcolorCounter
    \node[nodec, fill=currentcolor] at (2-1-h3) {};
    \node[nodec, fill=currentcolor] at (2-2-v2) {};
    
    \drawCrossFromToCC{2}{2}{3}{4}{1}{3}{1}{3}
    
    \setcolorCounter
    \node[nodec, fill=currentcolor] at (2-2-v4) {};
    \node[nodec, fill=currentcolor] at (2-3-h1) {};

    \setcolorCounter
    \node[nodec, fill=currentcolor] at (2-1-h2) {};
    \node[nodec, fill=currentcolor] at (3-2-v1) {};
    
    \setcolorCounter
    \node[nodec, fill=currentcolor] at (3-2-v2) {};
    \node[nodec, fill=currentcolor] at (2-2-h3) {};

    \setcolorCounter
    \node[nodec, fill=currentcolor] at (3-2-v3) {};
    \node[nodec, fill=currentcolor] at (2-3-h2) {};
        
    \setcolorCounter
    \node[nodec, fill=currentcolor] at (3-2-v4) {};
    \node[nodec, fill=currentcolor] at (2-2-h1) {};

\end{tikzpicture}
\end{externalize}}}$
    \caption[Example for an embedded subgraph structure]%
            {Example for an embedded subgraph structure of a single vertex with all outer neighbors, 
             extracted from the complete graph embedding in the broken Chimera graph of \autoref{fig:embedding}}
    \label{fig:subgraph}
\end{figure}
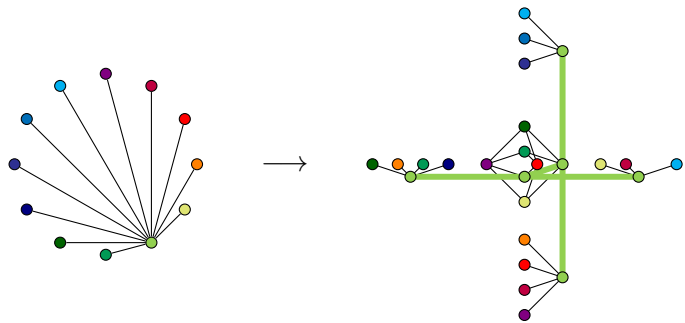

An example is shown in \autoref{fig:subgraph}. 
Note that the quadratic terms for the edges between the outer neighbors of the last point do not include variables corresponding to vertices in $\vertices$. 
Therefore, they are not considered in the definition of $\closure{I}\vphantom{I}^v$.  
In the following section, we nevertheless argue why we can omit these edges. 

Although, apart from the constraints
\begin{equation}
    S_{vw} = \sum_{pq \in \delta_{vw}} \closure{S}_{pq} \qquad \forall w \in N(v)
\end{equation} 
as stated in \autoref{sec:synch},
we are free to choose the values for the strengths on the outer edges $\closure{S}_{pq}$ for all $pq \in \outeredges$, 
this introduces another level of complexity to the overall problem. 
Their choice does not only concern the evaluated vertex $v$ but also its neighbors in $G$. 
We keep this additional level for future research and assume in the following that $\closure{S}_{pq}$ is validly chosen in advance and thus given and fixed. 
Nevertheless, we discuss possible choices supporting the simplification of the problem in the following section 
and see that our approach can be applied in any case. 

To differ between the outer edges with the given strength and the inner edges with the strength to be found, 
we rename them with $\pmb{\alpha} \vcentcolon = \closure{S}_{\closure{D}} \in \R^{\closure{D}}$ and $\pmb{\beta} \vcentcolon = - \closure{S}_{\closure{E}} \in \R^{\closure{E}}$.  
Note the extraction of the negative sign in the definition of $\beta$. 
For simplicity of the notation, we also use for the variable weights $\pmb{\omega} \vcentcolon = \closure{W}_{\varphi_v} \in \R^\vertices$.
With these notations and the simplifications of the previous section, we have for $r \in \{-1, 1\}^\vertices$ and $s \in \{-1, 1\}^\outerneighbors$
\begin{align}
    \closure{I}\vphantom{I}^v(r,s) %
    &= \sum_{q \in \vertices} \closure{W}_q r_q + \sum_{pq \in \edges} \closure{S}_{pq} r_p r_q + \sum_{q \in \vertices} \sum_{qn \in \outeredges} \closure{S}_{qn} r_q s_n \\
    &= \sum_{q \in \vertices} \omega_q r_q - \sum_{pq \in \edges} \beta_{pq} r_p r_q + \sum_{q \in \vertices} \sum_{qn \in \outeredges} \alpha_{qn} r_q s_n \\
    &= \vcentcolon \pmb{I_{\omega, \beta}^{\alpha}(r, s)},
\end{align}
the Ising model $I_{\omega, \beta}^{\alpha} : \{-1,1\}^\vertices \times \{-1,1\}^\outerneighbors \to \R$, 
where we drop the bar and the superscript $v$ for simplicity. 
All in all, we assume to be given 
\begin{itemize} 
    \item the strengths on the outer edges $\alpha \in \R^\outeredges$ and 
    \item the total weight $\pmb{\totalweight} \vcentcolon = W_v \in \R$
\end{itemize} 
and search for  
\begin{itemize}
    \item the weights $\omega \in \R^\vertices$ and
    \item the strengths on the inner edges $\beta \in \R^\edges$. 
\end{itemize} 

Note that we could `cut off' vertices from the embedded graph, where there are no outer edges incident to these vertices or all of them have zero strength, 
that means there is no `influence from outside' on these vertices. 
In turn, we thus assume we always have at least one outer edge with a non-zero strength, that is, $\{n \in \outerneighbors : \ell n \in \outeredges, \alpha_{\ell n} \neq 0\} \neq \emptyset$ for all leaves $\ell$ of $H[\varphi_v]$. 

\subsubsection{Output and Objective}

The larger we choose $\beta_{pq}$, the stronger the vertices $p$ and $q$ are coupled due to the negative sign in $I_{\omega, \beta}^{\alpha}$. 
As discussed in \autoref{sec:ising}, we cannot simply set these strengths to some very large value compared to the remaining parameters due to the machine restrictions.
In literature, the coupling strength is mentioned to be decisive for the success probability, 
however usually yielding the maximal absolute coefficient $C_{\max}$ at the same time. 
Therefore, the question arises: How small can we set these strengths such that we can still achieve an equivalent embedded Ising? 
This means that a first step based on current practice would be to simply minimize~$\|\beta\|_{\infty}$.
 
However, by only minimizing the strengths, a corresponding suitable weighting could exceed the corresponding bound in some vertices.  
Hence, with the strengths on the outer edges assumed to be fixed, 
the more interesting objective would be the maximal absolute value of all remaining parameters of the observed part of the Ising model $\max\{\|\omega\|_{\infty}, \|\beta\|_{\infty}\}$, 
which should be minimized in total.

Besides, by excluding possibly existing additional inner edges, $H[\varphi_v]$ could be reduced to a tree. 
We use this fact later on, as several problems are much easier on trees. 
However, for the first steps, we consider $H[\varphi_v]$ to be an arbitrary graph. 

\subsubsection{Constraints}

By the previous section, we can already derive the following constraints on the introduced parameters $\omega$ and $\beta$: 
The weights should sum up to the total weight with
\begin{equation}
    \totalweight = \sum_{q \in \vertices} \omega_q =\vcentcolon \pmb{\omega(\vertices)}
\end{equation} 

and, from the conditions of \autoref{lem:single_ok}, we need  
\begin{equation}\label{eq:constraint}
\begin{aligned}
    I_{\omega, \beta}^{\alpha}(r, s) > \min \left\{I_{\omega, \beta}^{\alpha}(-\One,s), I_{\omega, \beta}^{\alpha}(\One, s) \right\} 
        \quad&\forall s \in \{-1,1\}^\outerneighbors ~\ifjournalversion{\\ &}{}\forall r \in \{-1,1\}^\vertices \setminus \{-\One, \One\}, 
\end{aligned}
\end{equation}
to ensure that the full embedded problem is provably equivalent to original one in the end.

Note that the latter condition is comprised of an exponential number of constraints, more precisely $2^{|\outerneighbors|} \big( 2^{|\vertices|} - 2 \big)$ many.
Although they are linear inequalities, the overall optimization problem is therefore not solvable in polynomial time in a straightforward way. 
Thus, it could only be used for small embedded subgraph instances~$H[\varphi_v]$ in practice.   

By introducing a \emph{gap} value $\pmb{\gamma} \in \R_{> 0}$, we can further influence how `far away', in terms of the distance of their objective values, invalid variable assignments are from the valid ones.
This value might become important for the user of the D-Wave machine, when trying to improve the success probability of finding an optimal solution.  
By this we can also relax the order relation to a greater or equal: 
\begin{equation}
\begin{aligned}
    I_{\omega, \beta}^{\alpha}(r, s) \geq \min \left\{I_{\omega, \beta}^{\alpha}(-\One,s), I_{\omega, \beta}^{\alpha}(\One, s) \right\} + \gamma 
        \quad&\forall s \in \{-1,1\}^\outerneighbors ~\ifjournalversion{\\ &}{}\forall r \in \{-1,1\}^\vertices \setminus \{-\One, \One\}.
\end{aligned}
\end{equation}
In the following, we assume this value is given with the input and fixed. 
In future research, we might also investigate different approaches in trading off the gap against the strength. 
This however shall not be part of this work. 

\subsubsection{In a Nutshell}

By the previous notations and reformulations, we can now summarize the problem. 
While in~\cite{lobe2022diss} the problem was split up in two problems with two different objective functions,  
this is not relevant for our hardness result here.
Therefore we concentrate on the full problem with 

\begin{fprob*}[\textsc{Gapped Parameter Setting Problem}]\label{prob:original}
    \ Given an embedded subgraph structure $({\vertices \cupdot \outerneighbors}, \edges \cupdot \outeredges)$, $\alpha \in \R^\outeredges$, $\totalweight \in \R$ and $\gamma \in \R_{> 0}$, 
    find $\omega$ and $\beta$ that solve
    \begin{align}
        \min~        &\max \big\{ \|\omega\|_\infty, \|\beta\|_\infty \big\}\\
        \text{s.t. } &\omega \in \R^\vertices,\, \beta \in \R^\edges, \\
                     &\omega(\vertices) = \totalweight, \\
                     &I_{\omega, \beta}^{\alpha}(r, s) \geq \min \left\{I_{\omega, \beta}^{\alpha}(-\One, s), I_{\omega, \beta}^{\alpha}(\One, s) \right\} + \gamma\! &&\forall s \in \{-1,1\}^\outerneighbors\\ 
                        &&&\forall r \in \{-1,1\}^\vertices \setminus \{-\One, \One\}. \ifjournalversion{\\[-2\baselineskip]}{}
    \end{align}
\end{fprob*}

\subsection{Simplifications}\label{sec:simple}

The instance defined in the previous section can be simplified due to some properties of the Ising models. 
We can apply several steps, which are discussed in the following. 

\inputFromHere{simplifications/strength}

\inputFromHere{simplifications/symmetry}

\inputFromHere{simplifications/independent}

\inputFromHere{simplifications/single}

\inputFromHere{simplifications/cuts}

\inputFromHere{simplifications/summary}

\section{Analysis}\label{sec:analysis}

As $\delta(S)$ is constant for every set $S$, we only have linear functions and \refProb{both} thus belongs to the class of linear optimization problems (LPs).  
Although we have already reduced and simplified the requirements on the embedded Ising problem, 
we still have to take every possible constellation of the signs of the outer influences on the embedded vertices into account.
This is results in a constraint for every non-trivial subset of the vertices of $H[\varphi_v]$.
Their number is exponentially large and we therefore have exponentially large LPs, 
which cannot be solved in polynomial time in a straightforward way.
We need to analyse the problem in more detail.

\subsection{Polyhedral Description}\label{sec:poly}

In the following, let the graph $\pmb{G} = (\pmb{V}, \pmb{E})$, the strengths $\pmb{\sigma} \in \R_{\geq 0}^V$, 
the total weight $\pmb{\totalweight} \in \R_{\geq 0}$ with $\totalweight < \sigma(V)$ and the gap $\pmb{\gamma} \in \R_{> 0}$ be given and fixed. 
For $\emptyset \neq S_1 \subseteq S_2 \subseteq V$, we see the $\sigma$-sums are monotonic over the partial ordering of the subset relation with
\begin{equation}
    0 \leq \sigma(S_1) \leq \sigma(S_2) \leq \sigma(V)
\end{equation}
due to $\sigma \geq \Zero$. 
Furthermore, we have 
\begin{equation}
    \omega(S) + \omega(V \setminus S) = \omega(V) = \totalweight
\end{equation}
for arbitrary $\emptyset \neq S \subseteq V$. 
With
\begin{equation}
\begin{alignedat}{2}
    &&\sigma(S) + \omega(S) &\lesseqgtr \sigma(V \setminus S) - \omega(V \setminus S) \\
    &\Leftrightarrow &\sigma(S) &\lesseqgtr \tfrac{1}{2}\left(\sigma(V) - \totalweight\right) \\
    &\Leftrightarrow &\sigma(V \setminus S) &\gtreqless \tfrac{1}{2}\left(\sigma(V) + \totalweight\right) %
\end{alignedat}
\end{equation}
and 
\begin{equation}
     \sigma(V) + \totalweight \geq \sigma(V) \geq \sigma(V) - \totalweight
\end{equation}
for $\totalweight \geq 0$, we get by the resolution of the minimum the constraint
\begin{equation}
    \vartheta \geq \begin{cases}
        \frac{\sigma(S) + \omega(S) + \gamma}{|\delta(S)|} &\text{if } \sigma(S) < \tfrac{1}{2}\left(\sigma(V) - \totalweight\right), \\[1ex]
        \frac{\sigma(V \setminus S) - \omega(V \setminus S) + \gamma}{|\delta(S)|} &\text{otherwise}
    \end{cases}
\end{equation} 
for all $\emptyset \neq S \subsetneq V$.
This can be used for the following polyhedra and other helpful definitions:

\begin{definition}\label{def:thetaHalf} %
    Let 
    \begin{align}
        \pmb{\varthetaHalf} &\vcentcolon = \tfrac{1}{2} \left(\sigma(V) - \totalweight\right), \\
        \pmb{\cutpoly(S)} &\vcentcolon= \big\{ (\vartheta, \omega) \in \R \times \R^V: \vartheta|\delta(S)| \geq \min \{\sigma(S) + \omega(S), \sigma(V \setminus S) - \omega(V \setminus S)\} + \gamma \big\}, \\
        \pmb{\Theta_\totalweight}  &\vcentcolon= \big\{ (\vartheta, \omega) \in \R \times \R^V: \omega(V) = \totalweight \big\}, \\
        \pmb{\cutpoly}    &\vcentcolon= \bigcap_{\emptyset \neq S \subsetneq V} \cutpoly(S) \cap \Theta_\totalweight, \\
        \pmb{\Phi}      &\vcentcolon = \big\{(\vartheta, \omega) \in \R \times \R^V : \vartheta \geq \|\omega\|_\infty\big\}.
    \end{align}
\end{definition}
We have
\begin{equation}
    \pmb{\cutpoly(S)} \vcentcolon= \begin{dcases*}
                            \big\{(\vartheta, \omega) \in \R \times \R^V: \vartheta|\delta(S)| \geq \sigma(S) + \omega(S)  + \gamma \big\} &\normalfont if $\sigma(S) < \varthetaHalf$, \\
                            \big\{(\vartheta, \omega) \in \R \times \R^V: \vartheta|\delta(S)| \geq 2 \varthetaHalf - \sigma(S) + \omega(S) + \gamma \big\} &\normalfont otherwise
                        \end{dcases*}
\end{equation}
by the given relations and due to 
\begin{align}
    \sigma(V \setminus S) - \omega(V \setminus S) &= \sigma(V) - \sigma(S) - \omega(V) + \omega(S) \\
        &= \sigma(V) - \totalweight - \sigma(S) + \omega(S) \\
        &= \tfrac{1}{2} \varthetaHalf - \sigma(S) + \omega(S).
\end{align}
Note that, with $\totalweight < \sigma(V)$, we always have $\varthetaHalf > 0$. 

Now we can reformulate the problem: 
\begin{corollary}
    The \refProb[]{both} can be written as the LP
    \begin{equation}
        \min \left\{\vartheta : (\vartheta, \omega) \in \cutpoly \cap \Phi \right\}.
    \end{equation}
\end{corollary}

With these definitions, we can easily see that $\cutpoly$ is an unbounded, $|V|$-di\-men\-sion\-al polyhedron described by an exponential number of inequalities.
The domain of \refProb{both} is then the intersection of $\cutpoly$ with the cone $\Phi$, which is defined by only $2|V|$ constraints
since $\vartheta \geq \|\omega\|_\infty$ is equivalent to
\begin{equation}
    \vartheta \geq |\omega_v| \quad \forall v \in V,
\end{equation}
or without any absolute value also to 
\begin{equation}
\begin{aligned}
    \vartheta &\geq \omega_v && \forall v \in V, \\
    \vartheta &\geq - \omega_v && \forall v \in V.
\end{aligned}
\end{equation}

\subsection{Connected Vertex Sets}\label{sec:connected}

Note that we restrict our considerations to $0 \leq \totalweight < \sigma(V)$ to focus on the cases that are not preprocessable according to \autoref{lem:preprocess}.  
This implies $\varthetaHalf > 0$, which is indeed necessary for the constructions. 
We further assume $\pmb{G}$ and $\pmb{\sigma} \in \R^V_{\geq 0}$ to be given and fixed again. 

To reduce the complexity of the description of $\cutpoly$ from \autoref{def:thetaHalf}, we need to reduce the number of inequalities. 
By the following result, we can indeed show that some inequalities describing $\cutpoly$ are redundant due to the monotonicity of the $\sigma$-sums: 

\begin{theorem}\label{thm:connectedcutssuffice}
    With $\pmb{\fC} \vcentcolon = \big\{\emptyset \neq S \subsetneq V : \text{\normalfont $G[S]$ connected and $G[V \setminus S]$ connected}\big\}$, we have
    \begin{equation}
        \cutpoly = \bigcap_{S \in \fC} \cutpoly(S) \cap \Theta_\totalweight.
    \end{equation}
\end{theorem}

\begin{proof}
    Due to the reduction of the number of constraints, the left-hand side is contained in the right-hand side.
    In the following, we show the reverse direction by establishing the redundancy of the constraints associated with sets in $2^V \setminus \{\emptyset, V\}$ that are not included in $\fC$.
    For this we introduce the type $t(S)$ of a vertex set $\emptyset \neq S \subsetneq V$ denoting the number of connected components of the corresponding induced subgraph $G[S]$.
    Note that every non-empty vertex set has at least a type of 1 and we only have $t(S) = 1$ if $G[S]$ is connected. 
    As the equality $\omega(V) = \totalweight$ is important for the following derivations, 
    we furthermore work on the hyper\-plane~$\Theta_\totalweight$ but do not explicitly mention it for simplicity.
     
    The first step is to show the sufficiency of the restriction to vertex sets that induce connected subgraphs, which means 
    \begin{equation}
        \bigcap_{\emptyset \neq S \subsetneq V} \cutpoly(S) \supseteq \bigcap_{S \in \fC'} \cutpoly(S)
    \end{equation}
    with $\pmb{\fC'} = \big\{\emptyset \neq S \subsetneq V : \text{\normalfont $G[S]$ connected}\big\} = \big\{\emptyset \neq S \subsetneq V : t(S) = 1\big\}$.
    Suppose this does not hold. 
    Let $S^* \in 2^V \setminus (\fC' \cup \{\emptyset, V\}) = \big\{\emptyset \neq S \subsetneq V : t(S) > 1\big\}$ be a vertex set with 
    \begin{equation}\label{eq:contradict}
        \cutpoly(S^*) \not\supseteq \bigcap_{S \in \fC'} \cutpoly(S) %
    \end{equation}
    having minimal type. 
    Then there exist two non-empty vertex sets $S_1$ and $S_2$ with $S^* = S_1 \cupdot S_2$ and $\delta(S_1, S_2) = \delta(S_1) \cap \delta(S_2) = \emptyset$.
    Due to $t(S_1) + t(S_2) = t(S^*)$ and the minimality of $S^*$, we have $t(S_1) = t(S_2) = 1$
    and therefore $S_1, S_2 \in \fC'$ with   
    \begin{equation}\label{eq:supset}
        \cutpoly(S_1) \cap \cutpoly(S_2) \supseteq \bigcap_{S \in \fC'} \cutpoly(S). %
    \end{equation}

    In the following, we show that the inequality defining $\cutpoly(S^*)$ can be derived 
    from the inequalities defining $\cutpoly(S_1)$ and $\cutpoly(S_2)$ by summation. 
    Because
    \begin{equation}
        \{x \in X : f(x) + g(x) \leq 0\} \supseteq \{x \in X : f(x) \leq 0, g(x) \leq 0\} 
    \end{equation}
    holds for arbitrary domains $X$ and functions $f,g : X \to \R$,
    this results in 
    \begin{equation}\label{eq:summing}
        \cutpoly(S^*) \supseteq \cutpoly(S_1) \cap \cutpoly(S_2),
    \end{equation} 
    which is, together with \eqref{eq:supset}, a contradiction to \eqref{eq:contradict}.
    This is supported by the additivity of $\sigma$ and~$\omega$ with 
    \begin{equation}
    \begin{aligned}
        \sigma(S^*) &= \sigma(S_1) + \sigma(S_2), \\
        \omega(S^*) &= \omega(S_1) + \omega(S_2)
    \end{aligned}
    \end{equation} 
    and 
    \begin{equation}
        |\delta(S^*)| = |\delta(V \setminus S^*)| = |\delta(S_1)| + |\delta(S_2)| - 2|\delta(S_1, S_2)| = |\delta(S_1)| + |\delta(S_2)|. 
    \end{equation}
    
    We have two different cases concerning $S^*$: 
    \begin{enumerate}[A), leftmargin=*] 
        \item\label{item:first} $\sigma(S^*) < \varthetaHalf$: 
            Due to the monotonicity of $\sigma$, we also have $\sigma(S_1), \sigma(S_2) \leq \sigma(S^*) < \varthetaHalf$.
            From the inequalities defining $\cutpoly(S_i)$,
            \begin{equation}
                \vartheta |\delta(S_i)| \geq \sigma(S_i) +  \omega(S_i) + \gamma
            \end{equation}
            for $i=1,2$, 
            we get
            \begin{equation}
            \begin{aligned}
                \vartheta |\delta(S^*)| &= \vartheta |\delta(S_1)| + \vartheta |\delta(S_2)| \\ 
                    &\geq \sigma(S_1) +  \omega(S_1) + \sigma(S_2) +  \omega(S_2) + 2\gamma\\
                    &= \sigma(S^*) + \omega(S^*) + 2\gamma \\ 
                    &\geq \sigma(S^*) + \omega(S^*) + \gamma
            \end{aligned}
            \end{equation}
        with $\gamma > 0$, which provides the constraint defining $\cutpoly(S^*)$.  
        \item $\sigma(S^*) \geq \varthetaHalf$: 
        In this case, we cannot derive conditions on $\sigma(S_1)$ and $\sigma(S_2)$. 
        Thus, there are three possibilities, which follow the same construction as in case \ref{item:first}:
        \begin{enumerate}[a), leftmargin=*]
            \item $\sigma(S_1), \sigma(S_2) < \varthetaHalf$: From
                \begin{equation}
                    \vartheta |\delta(S_i)| \geq \sigma(S_i) +  \omega(S_i) + \gamma
                \end{equation}
                for $i=1,2$, we get
                \begin{align}
                    \vartheta |\delta(S^*)| &= \vartheta |\delta(S_1)| + \vartheta |\delta(S_2)| \\ 
                        &\geq \sigma(S_1) +  \omega(S_1) + \sigma(S_2) +  \omega(S_2) + 2\gamma\\
                        &\geq \sigma(S^*) + \omega(S^*) + \gamma
                \intertext{analogously to \ref{item:first} and further by `adding 0'}
                        &= 2\sigma(S^*) - \sigma(S^*) + \omega(S^*) + \gamma \\
                        &\geq 2\varthetaHalf - \sigma(S^*) + \omega(S^*) + \gamma.
                \end{align}
                Thus, we obtained the inequality of $\cutpoly(S^*)$ in this case.
            \item $\sigma(S_1), \sigma(S_2) \geq \varthetaHalf$: From
                \begin{equation}
                    \vartheta |\delta(S_i)| \geq 2\varthetaHalf - \sigma(S_i) + \omega(S_i) + \gamma
                \end{equation}
                for $i=1,2$ and $\varthetaHalf > 0$ for $\totalweight < \sigma(V)$, we get
                \begin{align}
                    \vartheta |\delta(S^*)| &= \vartheta |\delta(S_1)| + \vartheta |\delta(S_2)| \\ 
                        &\geq 4\varthetaHalf - \sigma(S_1) + \omega(S_1) - \sigma(S_2) + \omega(S_2) + 2\gamma \\
                        &\geq 2\varthetaHalf - \sigma(S^*) + \omega(S^*) + \gamma.
                \end{align}
            \item $\sigma(S_1) < \varthetaHalf$ and $\sigma(S_2) \geq \varthetaHalf$ w.l.o.g.: From
                \begin{equation}
                \begin{aligned}
                    \vartheta |\delta(S_1)| &\geq \sigma(S_1) +  \omega(S_1) + \gamma, \\
                    \vartheta |\delta(S_2)| &\geq 2\varthetaHalf - \sigma(S_2) + \omega(S_2) + \gamma,
                \end{aligned}
                \end{equation}
                we get
                \begin{equation}
                \begin{aligned}
                    \vartheta |\delta(S^*)| &= \vartheta |\delta(S_1)| + \vartheta |\delta(S_2)| \\
                        &\geq 2\varthetaHalf + \sigma(S_1) +  \omega(S_1) - \sigma(S_2) + \omega(S_2) + 2\gamma\\
                        &\geq 2\varthetaHalf - \sigma(S_1) +  \omega(S_1) - \sigma(S_2) + \omega(S_2) + 2\gamma\\
                        &\geq 2\varthetaHalf - \sigma(S^*) + \omega(S^*) + \gamma
                \end{aligned}
                \end{equation}
                due to $\sigma(S_1) \geq 0 \geq - \sigma(S_1)$.
        \end{enumerate}
    \end{enumerate}
    Therefore, the constraint associated with $S^*$ is redundant if $S^*$ does not induce a connected subgraph.
    
    Next we show that the vertex sets whose complement also does not induce a connected subgraph are unnecessary, too, 
    hence   
    \begin{equation}
        \bigcap_{S \in \fC'} \cutpoly(S) \supseteq \bigcap_{S \in \fC} \cutpoly(S).
    \end{equation}
    Note that, with the former definition of $\fC'$, we have $\fC = \big\{S \in \fC' : t(V \setminus S) = 1\big\}$.
    We derive an analogous contradiction as before and 
    therefore assume the above relation does not hold. 
    Let $S^* \in \fC' \setminus \fC = \big\{\emptyset \neq S \subsetneq V : t(S) = 1, t(V \setminus S) > 1\big\}$ be a vertex set with 
    \begin{equation}\label{eq:contra}
        \cutpoly(S^*) \not\supseteq \bigcap_{S \in \fC} \cutpoly(S), %
    \end{equation}
    whose complement $V \setminus S^*$ has minimal type. 
    We can derive analogously to the first part that $G[V \setminus S^*]$ needs to consist of only two connected components induced by vertex sets in~$\fC$. 
    However, we need a minor additional step:
    Let 
    \begin{equation}
        V \setminus S^* = \bigcupdot_{i=1}^k X_i,
    \end{equation}
    with $t(X_i) = 1$ for all $i=1, ..., k$ and $|\delta(X_i, X_j)| = \emptyset$ for all $i \neq j \in \{1, ..., k\}$, 
    be a partition of $V \setminus S^*$ into the vertex sets inducing the connected components of $G[V \setminus S^*]$ for $t(V \setminus S^*) = k > 1$.
    Since the graph~$G$ is connected, the complement of two of the above vertex sets, w.l.o.g.~$X_1$ and $X_2$, induces a graph, $G[V \setminus (X_1 \cup X_2)]$, 
    which needs to be connected, too. 
    Then we have $t(V \setminus (X_1 \cup X_2)) = 2$, which would contradict the choice of $S^*$ if we had $k>2$.
    This means we have $ V \setminus S^* = X_1 \cupdot X_2$ with $X_1, X_2 \in \fC$ and symmetrically also $V \setminus X_1, V \setminus X_2 \in \fC$.
    This is illustrated in \autoref{fig:connected}.
    
    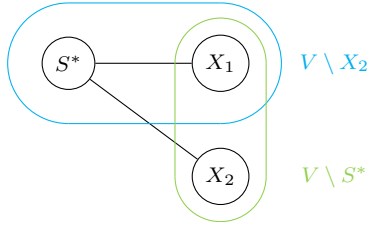
\begin{figure}[t!]
        \centering
        \newcommand{\drawAround}[4]{
	\path[name path=c1] (#1) circle[radius=#3];
	\path[name path=c2] (#2) circle[radius=#3];
	\path[name path=l1] let \p1=($(#1) - (#2)$) in ($(#1) + (\y1, -\x1)$) -- ($(#1) + (-\y1, \x1)$);
	\path[name path=l2] let \p1=($(#1) - (#2)$) in ($(#2) + (\y1, -\x1)$) -- ($(#2) + (-\y1, \x1)$);
	\path[name intersections={of=c1 and l1, by={r1, r2}}];
	\path[name intersections={of=c2 and l2, by={r3, r4}}];
	\draw[#4] (r1) -- (r3);
	\draw[#4] (r2) -- (r4);
	\path[draw, #4] let \p1=($(#2) - (r3)$) in pic[draw,angle radius={veclen(\x1,\y1)}] {angle = r4--#2--r3};
	\path[draw, #4] let \p1=($(#1) - (r1)$) in pic[draw,angle radius={veclen(\x1,\y1)}] {angle = r1--#1--r2};
}

\newcommand{\drawAroundThree}[5]{
	\path[name path=c1] (#1) circle[radius=#4];
	\path[name path=c2] (#2) circle[radius=#4];
	\path[name path=c3] (#3) circle[radius=#4];
	\path[name path=l12] let \p1=($(#1) - (#2)$) in ($(#1) + (\y1, -\x1)$) -- ($(#1) + (-\y1, \x1)$);
	\path[name path=l21] let \p1=($(#1) - (#2)$) in ($(#2) + (\y1, -\x1)$) -- ($(#2) + (-\y1, \x1)$);
	\path[name path=l13] let \p1=($(#1) - (#3)$) in ($(#1) + (\y1, -\x1)$) -- ($(#1) + (-\y1, \x1)$);
	\path[name path=l31] let \p1=($(#1) - (#3)$) in ($(#3) + (\y1, -\x1)$) -- ($(#3) + (-\y1, \x1)$);
	\path[name path=l23] let \p1=($(#2) - (#3)$) in ($(#2) + (\y1, -\x1)$) -- ($(#2) + (-\y1, \x1)$);
	\path[name path=l32] let \p1=($(#2) - (#3)$) in ($(#3) + (\y1, -\x1)$) -- ($(#3) + (-\y1, \x1)$);
	
	\path[name intersections={of=c1 and l12, by={r1, r2}}];
	\path[name intersections={of=c2 and l21, by={r3, r4}}];
	\path[name intersections={of=c1 and l13, by={s1, s2}}];
	\path[name intersections={of=c3 and l31, by={s3, s4}}];
	\path[name intersections={of=c2 and l23, by={t1, t2}}];
	\path[name intersections={of=c3 and l32, by={t3, t4}}];
	\draw[#5] (r1) -- (r3);
	\draw[#5] (s2) -- (s4);
	\draw[#5] (t1) -- (t3);
	\path[draw, #5] let \p1=($(#1) - (r1)$) in pic[draw,angle radius={veclen(\x1,\y1)}] {angle = r1--#1--s2};
	\path[draw, #5] let \p1=($(#2) - (r3)$) in pic[draw,angle radius={veclen(\x1,\y1)}] {angle = t1--#2--r3};
	\path[draw, #5] let \p1=($(#3) - (t3)$) in pic[draw,angle radius={veclen(\x1,\y1)}] {angle = s4--#3--t3};
}

\begin{tikzpicture}

	\tikzstyle{set}=[circle, draw, fill=white, inner sep=3]

	\clip (-1, -2.1) rectangle (4.5, 0.8);
	
	\coordinate (S);
	\coordinate (X1) at ($(S) + (2, 0)$);
	\coordinate (X2) at ($(S) + (2, -1.5)$);
	
	\node[set] (S) at (S) {$S^*$};
	\node[set] (X1) at (X1) {$X_1$};
	\node[set] (X2) at (X2) {$X_2$};
    
    \draw (S) -- (X1); 
    \draw (S) -- (X2); 

	\drawAround{S}{X1}{0.8}{lightblue}
	\node at ($(X1) + (1.5,0)$) {\color{lightblue}$V \setminus X_2$};

    \drawAround{X1}{X2}{0.6}{lightgreen}
    \node at ($(X2) + (1.5,0)$) {\color{lightgreen}$V \setminus S^*$};
	
\end{tikzpicture}
        \caption{Connected components for unconnected $G[V \setminus S^*]$}
        \label{fig:connected}
    \end{figure}
    
    The construction follows the same structure as in the first part,  
    only with a slight deviation, because we use the complements in the following relations:  
    \begin{equation}\label{eq:relations}
        \cutpoly(S^*) \supseteq \cutpoly(V \setminus X_1) \cap \cutpoly(V \setminus X_2) \supseteq \bigcap_{S \in \fC} \cutpoly(S).
    \end{equation}
    In order to establish the desired contradiction to \eqref{eq:contra}, the first relation remains to be shown.
    Remember, due to the additivity, we have
    \begin{equation}
    \begin{aligned}
        \totalweight &= \omega(S^*) + \omega(X_1) + \omega(X_2), \\
        \sigma(V) &= \sigma(S^*) + \sigma(X_1) + \sigma(X_2), \\
        |\delta(S^*)| &= |\delta(V \setminus S^*)| = |\delta(X_1)| + |\delta(X_2)|.
    \end{aligned}
    \end{equation}
    
    Analogously to before, we need to distinguish between different cases. 
    However, the different possibilities for~$S^*$ `arise naturally' and we therefore only consider the sets $X_1$ and $X_2$:      
    \begin{enumerate}[a), leftmargin=*]
        \item $\sigma(V \setminus X_1), \sigma(V \setminus X_2) \geq \varthetaHalf$:
            From the inequalities defining $\cutpoly(V \setminus X_1)$ and $\cutpoly(V \setminus X_2)$, 
            \begin{equation}
                \vartheta |\delta(X_i)| = \vartheta |\delta(V \setminus X_i)| \geq 2\varthetaHalf - \sigma(V \setminus X_i) + \omega(V \setminus X_i) + \gamma
            \end{equation}
            for $i=1,2$, we get
            \begin{align}
                \vartheta |\delta(S^*)| &= \vartheta |\delta(X_1)| + \vartheta |\delta(X_2)| \\ 
                    &\geq 4\varthetaHalf - \sigma(V \setminus X_1) + \omega(V \setminus X_1) - \sigma(V \setminus X_2) + \omega(V \setminus X_2) + 2\gamma \\
                    &= 4\varthetaHalf - \sigma(V \setminus X_1) + \omega(V \setminus X_1) - \sigma(S^* \cup X_1) + \omega(S^* \cup X_1) + 2\gamma \\
                    &= 4\varthetaHalf - \sigma(V) + \sigma(X_1) + \totalweight - \omega(X_1) - \sigma(S^*) - \sigma(X_1) + \omega(S^*) + \omega(X_1) + 2\gamma \\
                    &\geq 2\varthetaHalf - \sigma(S^*) + \omega(S^*) + \gamma,
            \intertext{which is the inequality defining $\cutpoly(S^*)$ in case of $\sigma(S^*) \geq \varthetaHalf$.
                       For $\sigma(S^*) < \varthetaHalf$, we can extend the chain of relations by}
                    &> 2\sigma(S^*) - \sigma(S^*) + \omega(S^*) + \gamma \\
                    &= \sigma(S^*) + \omega(S^*) + \gamma,
            \end{align} 
            which provides the inequality defining $\cutpoly(S^*)$ in this case.
        \item $\sigma(V \setminus X_1), \sigma(V \setminus X_2) < \varthetaHalf$: 
            These conditions contradict each other due to 
            \begin{equation}
                \sigma(X_1) = \sigma(V) - \sigma(V \setminus X_1) 
                    > \sigma(V) - \tfrac{1}{2}\big(\sigma(V) - \totalweight\big) = \tfrac{1}{2}\big(\sigma(V) + \totalweight\big) \geq \varthetaHalf, 
            \end{equation}
            while we require at the same time $\sigma(V \setminus X_2) = \sigma(X_1) + \sigma(S^*) < \varthetaHalf$. 
        \item $\sigma(V \setminus X_1) < \varthetaHalf$ and $\sigma(V \setminus X_2) \geq \varthetaHalf$ w.l.o.g.:  
            From %
            \begin{equation}
            \begin{aligned}
                \vartheta |\delta(X_1)| = \vartheta |\delta(V \setminus X_1)| &\geq \sigma(V \setminus X_1) + \omega(V \setminus X_1) + \gamma \\
                \vartheta |\delta(X_2)| = \vartheta |\delta(V \setminus X_2)| &\geq 2\varthetaHalf - \sigma(V \setminus X_2) + \omega(V \setminus X_2) + \gamma
            \end{aligned}
            \end{equation}
            for $i=1,2$, we get with
            \begin{align}
                \vartheta |\delta(S^*)| &= \vartheta |\delta(X_1)| + \vartheta |\delta(X_2)| \\ 
                    &= 2\varthetaHalf + \sigma(V \setminus X_1) + \omega(V \setminus X_1) - \sigma(V \setminus X_2) + \omega(V \setminus X_2) + 2\gamma \\
                    &= 2\varthetaHalf + \sigma(S^* \cup X_2) + \omega(S^* \cup X_2) - \sigma(V \setminus X_2) + \omega(V \setminus X_2) + 2\gamma \\
                    &= 2\varthetaHalf + \sigma(S^*) + \sigma(X_2) + \omega(S^*) + \omega(X_2) - \sigma(V) + \sigma(X_2) + \totalweight - \omega(X_2) + 2\gamma \\
                    &= \sigma(S^*) + 2\sigma(X_2) + \omega(S^*) + 2\gamma \\
                    &\geq \sigma(S^*) + \omega(S^*) + \gamma 
            \intertext{the inequality defining $\cutpoly(S^*)$ in case of $\sigma(S^*) < \varthetaHalf$ or by extending}
                    &= 2\sigma(S^*) - \sigma(S^*) + \omega(S^*) + \gamma \\
                    &\geq 2\varthetaHalf - \sigma(S^*) + \omega(S^*) + \gamma
            \end{align}
            the inequality defining $\cutpoly(S^*)$ in the other case where $\sigma(S^*) \geq \varthetaHalf$.
    \end{enumerate}
    Therefore, the constraint for $S^*$ is also redundant if $V \setminus S^*$ is not connected. 
\end{proof}

Although the $\cutpoly$-polyhedron can now be described with less inequalities, 
their number might still be exponential in an arbitrary graph. 
By the requirements of an embedding however, we can always restrict ourselves to trees by simply ignoring surplus edges in the graph.
For a lot of problems, it is known that trees yield better solvability because of their much simpler structure.
This also holds in our case and we can finally state: 

\begin{theorem}
    If $G$ is a tree, \refProb{both} can be solved in polynomial time.
\end{theorem}
\begin{proof}
    For $G$ being a tree, it is easy to see that the set of cuts $\fC$ of \autoref{thm:connectedcutssuffice} is only formed by cutting at single edges of the tree.
    More formally, this means 
    \begin{equation}
        \fC = \bigcup_{e \in E(G)}\{S_e, V \setminus S_e\},
    \end{equation}
    where $(S_e, V \setminus S_e)$ denotes the partitions derived from cutting the tree at edge $e$.
    Thus, we have $|\fC| = 2|E| = 2|V| - 2$.
    Together with the single inequality for $\Theta_\totalweight$ and the $2|V|$ additional inequalities in the full problem version describing $\Phi$, 
    the LP for \refProb{both} has a linear size in the number of vertices.
    Therefore, we can derive the stated result due to the general solvability of polynomial LPs.
\end{proof}
    
This means, we can use a standard LP solver to obtain a suitable weighting and the corresponding strength 
which are needed to formulate a provable equivalent \refProbNamed[]{embising} to a given one. 
In particular, the additional weight bounds do not increase the complexity of the problem.
Moreover, the thus found weights are optimized regarding the height of the final coefficients of the Ising problem 
and therefore promise a better performance of the quantum annealing machines.

\section{Conclusion}\label{sec:conclusion}

Whether the quantum annealers, in particular those built by D-Wave, 
show an advantage over classical computers is still under discussion 
and will only reveal with the further development of such machines. 
To evaluate the great potential of this technology however, 
we need to design our experiments carefully. 
If the architecture does not change drastically, 
the two programming steps, minor embedding and parameter setting, will remain relevant in the long term.  
They are critical when it comes to providing meaningful input for the annealing machines.  
While the necessary minor embedding can prevent calculations on the machine at all, that is, if no embedding can be found, 
the specific parameter setting decisively influences the success in solving the actual problem. %

We provide the first polynomial but rigorous description to find the parameters of the embedded Ising problem for a given problem and its corresponding embedding
such that both problems are provably equivalent. 
Due to the structure of the embeddings, the restriction on trees, required for the solvability proof,
is a condition which can always be established, e.g., by ignoring surplus edges in the embedded subgraphs.
Furthermore, while the embedding problem is strongly related to the specific hardware graphs of D-Wave, 
our weight distribution approach is applicable for all hardware which implements an Ising problem over graphs that do not yield an all-to-all connectivity and thus require an embedding.
Although our formulation makes several assumptions on the structure of the problem, 
\autoref{sec:simple} clearly shows where they come from and in which cases they might deviate from the actual bound. 
In turn we can deduce that there are several instances for which our bounds are tight.   

Thinking ahead, the parameter precision issues support the assumption that it is preferable to have as few different values for the parameters of the Ising problem as possible, 
while those should have the largest possible pairwise absolute difference, to achieve an acceptable success probability. 
If our original Ising problem is already defined only over integer parameters, 
a possible approach would be to also allow only integer parameters for the embedded Ising problem rather than dealing with rational parameters. 
This fixes the distance of the parameters to at least 1, which might increase the success probability.  
Minimizing the largest absolute integer parameter however increases the complexity of the problem significantly, as it is the step from an LP to an ILP.  
The evaluation of this problem will be the next step in our research. 

Another interesting research direction might be to extend the approach of \cite{raymond2020improving}: 
Rather than evaluating the worst case scenario, as we do in this publication, where the embedded vertices must be synchronized for all possible neighboring constellations, 
the average case, where they `only' must hold for most of the cases, might also suffice in practice but yield an even smaller coupling strength. 

Apart from that, with our results, the formulated Ising problems can now directly be transferred to the D-Wave machines. 
This means, the computational properties of these machines can now be investigated even further: 
Do the theoretically optimal coupling strengths also hold under the perturbations of the machine, 
that is, do they suffice to enforce the synchronization of the variables in practice?    
While, in theory, any positive gap is sufficient for the equivalence of the problems,
an ideal quantum annealer should thus indeed return the optimal solution after a sufficiently long annealing time,
the gap value most likely needs to be increased for any physical annealing machine. 
This gap parameter is a tool which provides different but, most importantly, equivalent encodings of the same problem
and thus allows to study the difference between the theoretical and the \emph{effective} coupling strength, 
which means the one which is necessary for the real machine to return the optimal value with an acceptable success probability.
With this we will get a deeper understanding of the problem-independent behaviour of such machines, 
supporting their further development. %

    \printbibliography
    
\end{document}